\documentstyle[12pt,aaspp4]{article}
\begin{document}
\def\be{\begin{equation}}
\def\ee{\end{equation}}
\def\ba{\begin{eqnarray}}
\def\ea{\end{eqnarray}}
\newlength{\royalength}
\setlength{\royalength}{6.5 true in}

\title{\Large Formation and Cosmic Evolution of Elliptical Galaxies}
\author{
{\bf J.A. de Freitas Pacheco$^1$,  R. Michard$^2$,  R. Mohayaee$^3$ }} 
\affil{Departments Fresnel$^1$ and
Cassini $^3$, Observatoire de la C\^ote d'Azur,
BP 4229, F-06304, Nice Cedex 4, France\\
$^2$Observatoire de Paris, LERMA, 77 Av. Denfert-Rochereau,
75015, Paris, France} 
\date {September 2002}

\vspace {1cm}

\begin{abstract}

A review of the present observational and theoretical status
of elliptical galaxies is presented with the aim to
clarify whether the monolithic or the
hierarchical, is a more viable scenario for the origin of these structures.
We describe the dynamical structure of elliptical galaxies 
using photometric and
spectroscopic data, in particular 3D observations from integral
field spectrographs, with emphasis on properties such as
brightness distribution, velocity profiles, central structures
like "cuspy" profiles and "cores", as well as central super massive black holes. 
We review also the main relations between these quantities like the fundamental 
plane and
colour-luminosity diagram. We present observational evidences for the 
presence of dark matter in the elliptical galaxies and discuss the
theories of galaxy formation within the framework of a 
cold dark matter cosmological model.
We discuss the formation of large-scale
structure, Press-Schechter theory and universal
density profile of dark matter halos. Subsequently, gas dynamics, star
formation, feedback, angular momentum, morphology and the epoch of galaxy
formation are studied and comparison between disks and spheroids are made.
Valuable insights into the formation epoch of elliptical galaxies
are provided by deriving the mean 
metallicities and ages at different redshifts through the study of
 different population synthesis models
(single-burst and evolutionary models) in particular the magnesium to iron 
ratio.
\end{abstract}

\vspace{2.0cm}
\noindent
To appear in "Recent Research Developments in Astronomy \& 
Astrophysics"

\newpage

\section{Introduction}

For many years after their introduction as one of the Hubble types, elliptical 
galaxies (E's) were 
generally thought of as ''boring dynamical systems", that is oblate spheroids 
supported by rotation. Unduly slow rotations however, were measured 
in a number of objects (Bertola \& Capaccioli 1975; Illingworth 1977). 
{\em The anisotropy of the velocity field was then invoked to
explain the flattening of ellipticals}.   
On the other hand a paper entitled ''A numerical model for a triaxial stellar 
system in dynamical equilibrium" was published by Schwarzschild (1979): it 
aimed at explaining the correlated radial variations in isophotes 
axis-ratios and orientations.  
Although these properties had already been well-documented in the sixties, 
the interest was renewed from both the theoretical and observational points of 
view (Williams \& Schwarzschild 1979a, 1979b; Leach 1981).    
{\em Triaxiality became fashionable}, and it was realized that 
ellipticals were quite exciting dynamical systems after all!

A few years later, the presence of disks in E's was advocated 
(Michard 1984, 1985), on the basis of the ellipticity radial profiles of E's 
as compared to S0 and the pointed forms of isophotes. A key result was 
published by Bender (1988), who found a correlation between 
isophotal geometry and kinematics: objects with pointed, or ''disky", 
isophotes are fast rotators and largely supported by rotation. 
Objects with ''boxy" isophotes rotate quite slowly, and can 
only be supported by ''thermal" motions, i.e. a stochastic system of 
stellar orbits. 
On average the fainter objects are disky ($di$E), while giant E's are often boxy 
($bo$E). A face-on disk cannot be readily detected, but is still 
present (Rix \& White 1990), hence the sub-class $un$E.  

Although most E-galaxies have a disk,  
this component is only a minor ornament, compared to the powerful rotating 
disks of spirals: in these galaxies the disks eventually dominates the 
morphology, photometry and kinematics. Spirals (S) have plenty of interstellar 
matter (ISM), including "cold gas" (atomic and molecular), dust and are 
accordingly 
still able to form new stars. Spirals are supported by rotation, 
including their central spheroidal bulge, often considered as a small 
E-galaxy ''grafted" upon the disk. In contrast, ellipticals have little 
ISM  and their stars are normally quite old. 
They have however the rather unique property to contain significant 
amounts of ''hot gas" at some $10^7$ K and therefore to emit X-rays. 
Rotation plays only a minor role in their dynamical support. 
Finally, radio-galaxies with extended lobes seem to be giant 
E's: is it a way to recognize these at large $z$?    

Of special interest are the lenticulars or S0 galaxies, intermediate between 
spirals and ellipticals: they have disks of various importance, between 
those of E and S objects, and are largely supported by rotation. No spiral 
structure appears but bar instability is current.  
On the other hand their stellar populations resemble those of ellipticals,  
while their amount of ISM varies very much from one object to the other: 
eventually as low as for E's, it may become quite noticeable, without allowing 
star formation.

It is a challenge for astrophysicists to understand the origin and evolution 
of  all these three major types of bright galaxies (not counting the various 
classes of dwarfs which will not be discussed in this review). Among the 
usual constraints are the relative populations of E/S0/S as a function 
of $z$ and in various environments, i.e. the field, small groups and 
clusters of diverse richness.    
We are nevertheless justified in considering separately the problem of 
ellipticals, because of a number of  systematic properties, in their  
photometry and kinematics, or in their stellar population content. 
These supply adequate guidance in deriving theoretical models, and precious 
tests of their success.

The origin of cosmic structures, including galaxies of all types, 
is currently described through the gravitational dissipationless 
collapse in a Universe dominated by Cold Dark Matter (CDM), with a suitable mix 
of primeval gas and photons. 
The existence of large amounts of dissipationless DM in 
the universe, clustered hierarchically as numerical simulations indicate, 
may induce the infall of gas into the potential wells of pre-existing halos 
and form the "luminous" galaxies. Note that the presently popular version 
of the CDM scenario, i.e. the $\Lambda$CDM (Bahcall et al. 1999) agrees with 
observations, such as the residual fluctuations in the cosmic 
microwave background (CMB) or the distribution of galaxy clusters and 
superclusters.    
It is of course unfortunate that the physical 
nature of CDM remains unknown, while the presence of DM halos around 
galaxies is still questioned from time to time, even for spirals  
which seemed to present the most robust dynamical proofs. 

Starting from preexisting DM halos (or protogalaxies) it is possible to model 
the formation of galaxies, using simulations which take into account the 
dynamics of DM and gas, star formation, radiative cooling and 
gas loss from galactic winds. Two main families of models may be recognized:  
\begin{enumerate}
\item The {\it monolithic scenario}, developed since the early work by 
Larson (1975), is characterized by an important star formation 
activity and a consequent chemical enrichment {/em at very early phases} 
of the galaxy history, followed by a long period of quiescent evolution. 
Episodes of star formation may occur later in the lifetime of 
the galaxy, stimulated by gas accreted from the surrounding ambient or by 
interactions with neighbors (Bertola et al. 1992). 

\item In the so called {\it hierarchical scenario}, 
ellipticals are formed from different merging episodes  through the Hubble 
time, which trigger the star formation activity and the chemical enrichment of 
the system (White \& Rees 1978).  
In this scenario, massive ellipticals form at relatively low
redshifts ($z \leq 1.5$) through the merging of spiral 
galaxies (Toomre 1977; Baugh et al. 1996, 1998). 
\end{enumerate}
As emphasized by Peebles (2002), the distinction between the two
scenarios is more of historical 
than physical significance. Large objects may be formed by interaction and 
merging 
of a hierarchy of primeval "lumps" in a protogalaxy: this is practically 
equivalent to a monolithic scenario. 
On the other hand, there is no doubt that
galaxies do interact, with sometimes spectacular effects, and have perhaps 
interacted still more at large z. The central problem is rather to 
find out when the bulk of the stars were formed and if this 
population evolved passively, or was modified, and to what extent, by 
interactions. Another important problem is the morphology: were E's formed with 
their present form, symmetry and lack of fast rotation (at least in the bright 
baryon dominated region), or did successive mergers transfer outwards the 
angular momentum of merging disks? 
Other relevant questions may result 
from the consideration of central super massive black holes (SMBH), now believed 
to be 
present in all stellar spheroids, and this from the "old days" of peak quasar 
activity.

The necessary constraints, or part of these, are given by the recognition 
of E's at larger and larger $z$ and their statistics in comparison with 
other Hubble types (if traceable). If formed at relatively low redshifts
(say $z \leq 1.5$), following a succession of mergers, E's should be rare 
or absent at distinctly larger $z$ values.  
In contrast, the ''monolithic' scenario predicts that massive ellipticals,  
completed at $z \approx 2-3$ after a brief and intense star formation period, 
are evolving passively, implying a constant comoving number density. 

A large body of observations show that ellipticals 
were already formed at $z \approx 2$ in about the same proportion as today, 
and have since then evolved passively. This results from studies of the 
Fundamental Plane, of the $z$ evolution of the color-luminosity 
relation or the Mg$_2-\sigma_d$ relation. Only minor events of 
''recent" stellar formation are allowed for perhaps 20\% of E's, but less in 
rich clusters. On the other hand, the oldest known galaxy is at $z=6.56$ 
(Hu et al., 2002), and a few have been discovered at $z > 5$. 
Quasars are known at similar $z$. This leaves a large gap of $z$ space to 
explore, for further progresses in the evolution of galaxies, including E's.

In Section 2 we review recent photometric and 
spectroscopic data leading to an improved picture of the rather complex 
dynamical structure of ellipticals. We present briefly the 
3D observations from Integral Field Spectrographs which are expected to 
bring major progresses when fully analyzed. We discuss the scaling relations 
such as the size-luminosity (or Kormendy's) relation and the Fundamental Plane, 
the color-luminosity or the Mg$_2-\sigma_d$ relations, which 
all have important impacts upon the problems of the evolution of E's. 
The empirical evidence for SMBH is also presented in this section.    

Section 3 introduces the evidence for the presence of Dark Matter halos 
around ellipticals, and then summarizes the CDM theory of the formation of 
cosmic structures, including protogalactic halos, leading to the birth  
of galaxies through gas concentration, stellar formation and feed-back to the 
gas from massive stars evolution. Gas dynamics and galaxy formation are 
discussed in Section 4. 

In Section 5, we discuss different population synthesis models
(single-burst and evolutionary models) used to interpret 
the integrated properties of stellar populations, and to derive mean 
metallicities and ages at different redshifts, thus testing the formation 
and evolutionary scenarios of E's, with the help of the precious clock, namely,
the [Mg/Fe] ratio,  provided by the theory of stellar evolution. 
 
Finally, in Section 6 we will  summarize the relevant observational and 
theoretical 
aspects concerning elliptical galaxies discussed in the present paper. 


\section{Observational constraints}

\subsection{Classical but basic photometry}

The essential parameters describing a galaxy are its mass M, its total 
luminosity L and a suitably defined ''radius" $r_e$. The ''natural" definition 
of  
$r_e$ is the so-called {\em effective radius, or de Vaucouleurs' radius}, 
defined from the two axis $a$ and $c$ 
of the (nearly elliptical) isophote encircling half the total light as
$r_e=(ac)^{1/2}$. A related quantity of interest is the average 
surface brightness $SB_e$ within the effective radius. 
The total magnitudes $m_T$, in one or the other of the usual pass-bands, are 
substitute to the luminosity. 

Photometric measurements need to be corrected for extinction by the 
interstellar Milky Way material. The corrections have long been derived 
from the HI line optical depth. New dust extinction maps have recently been 
produced from far IR satellite data (Schlegel et al. 1998).  
So-called K-corrections are needed to take care of the  
changes of the filter pass-band in the rest spectrum of 
the red-shifted objects. Cosmological corrections have to be introduced in 
the study of distant galaxies, beginning with $SB_e$ which decreases as 
$(1+z)^4$. For the so-called giant and intermediate 
ellipticals treated here, orders of magnitude are -22.8 to -18.6 for 
$m_T$ in B light, 25 to 1.5 kpc for $r_e$, $6\times 10^{10}$ L$_\odot$ to 
$1.6\times 10^9$ L$_\odot$ for L (in B light) and $2\times 10^{12}$ to $10^9$ 
M$_\odot$
for M.     


Distances are needed for the physical interpretation of photometric data. 
Clusters provide precious collections of galaxies at the same distance. 
In recent years tremendous progresses have been made in the evaluation of the
distances of E's, with the Fundamental Plane (see Section 2.6) and the 
technique of Surface Brightness Fluctuations (Tonry \& Schneider, 1988; 
Tonry et al., 2001). 
The results of both method are compared by Blakeslee et al. (2001).
 
It has long been known that the essential global photometric parameters, 
i.e. $m_T$ and $r_e$ (or $m_e$) are intercorrelated (Kormendy 1977; 
Michard 1979), suggesting some standard design of these systems. The 
name of ''Kormendy relation" recently appeared in the literature. This 
question has been revisited by Capaccioli et al. (1992). These statistics
 clearly indicate that dwarf ellipticals (dE) are not 
structurally similar to normal E.

The study of the radial surface-brightness profiles of E-galaxies is 
of great interest for a discussion of their evolution, because it shows rather
surprising regularities, which have been known for half a 
century (de Vaucouleurs  1948). The isophotal surface-brightness $\mu (r)$ 
(in magnitude) follows closely
the law $\mu (r)= \mu(r_e)+8.327[1-(r/r_e)^{1/4}]$, called 
de Vaucouleurs' law, and this in a large range of $\mu(r)$, more than 
10 magnitudes according to  de Vaucouleurs \& Capaccioli (1979). 
A substitute to the de Vaucouleurs' law, but of straightforward 
theoretical significance, has been given by Hernquist (1990). 
Systematic deviations from the de Vaucouleurs 
law have been noted. Kormendy (1977) introduced the concept of ''tidal halos" 
around large ellipticals with close neighbors, a paradoxical feature of 
Brightest Cluster Members (or cD galaxies). 
It has been also known that giant ellipticals often have an enhanced envelope 
above the $r^{1/4}$ law valid for their inner regions. Small ellipticals, 
seemingly satellites to much larger neighbors, show $\mu(r)$ profiles 
''truncated" outwards below the law valid for their inner regions. 
Altogether the deviations from the de Vaucouleurs law correlate with 
luminosities.  

These deviations have prompted a search for other representations 
of the $\mu(r)$ profiles. A review is due to Cappaccioli (1988).     
Systematic deviations from the $r^{1/4}$ law may be lessened by replacing the 
$1/4$ exponent by $1/n$ (S\'ersic 1968) with $n$ between 1 (exponential profile) 
and up to 10 or more. Applications of S\'ersic's formula can be found in 
Caon et al. (1993). 
As expected, the residuals are strongly reduced by adding one 
more free parameter in the fit! There is some correlation between the optimal 
$1/n$ values and the luminosity of the galaxies. 

In summary, the $r^{1/4}$ law is a remarkable property of {\em bona fide} 
ellipticals, but one cannot expect these products of Nature to comply rigidly 
with the rule! 
Systematic deviations exist, function of the luminosity. Local perturbations 
may result from the presence of embedded disks in the dominant $r^{1/4}$ 
spheroid. 

A variety of $\mu(r)$ profiles are presented in Figure 3. They have been 
obtained by merging HST (high resolution) and ground based (low resolution) 
data. Since the chosen abscissae 
scale is $(r/r_e)^{1/4}$, where $r_e$ is the measured de Vaucouleurs' radius, 
the graphs for various galaxies following the $r^{1/4}$ law would be parallel 
straight lines. One can appreciate from  Figure 3 the robustness of this 
law and the importance of the deviations from it. These are particularly noticeable 
towards the  center (see below). 


\subsection{Core and ''power-law" nuclei}

Nieto et al. (1991) already proposed to sort out E's into two subclasses 
according to their central photometric profiles: one group has relatively flat 
and 
broad {\em cores}, easily resolved from high resolution ground based 
observations.  
The other displays sharp peaked profiles, remaining unresolved 
under the same conditions. In the first group one finds {\em 
bright objects} of the $bo$E class (or $un$E) while the second contains the 
fainter $di$E objects, plus a variety of {\em not bright} $bo$E. These findings 
have been amply confirmed by the HST observations, already in its 
post-launching state 
(Jaffe et al. 1994; Lauer et al. 1995; Faber et al. 1997), and later after
installation of the correcting system (Carollo et al. 1997a, 1997b, 
Rest et al. 2001). 
The terminology of ''core-like" and ''power-law" profiles respectively are 
adopted here from Lauer et al. (1995). The HST profiles have been described by 
an ad hoc analytical law (''Nuker law") with no less than 5 
parameters (Byun et al. 1996). They have been 
de-projected to give the corresponding density profiles: the distinction 
between the two types is emphasized by the de-projection (Gebhardt et al. 1996). 

In  Figure 3, the two classes of central profiles are documented in the 
$(r/r_e)^{1/4}$ systems of abscissae. The differences 
are striking: the ''power-law" 
profiles are cusped above the $r^{1/4}$ law, while the ''core-like" profiles 
present a truncated $r^{1/4}$ law. 

The correlations between the two classes of nuclei and other galaxies properties  
(i.e. luminosity, disky/boxyness and rotation velocity) and their origins are  
discussed by Faber et al. (1997). 
Most explanations of the two kind of profiles invoke central black-holes 
(see Sect. 2.5). Silva \& Wise (1996) proposed however, that core-like nuclei 
could be ''flattened" by a central concentration of dust. This should produce 
a strong maximum of reddening, which is not present (Carollo et al. 1997a; 
Michard 1998c). 
It is rather the power-law nuclei which show a central red peak, of uncertain 
origin: dust concentration, or sharp metallicity increases or both. 

Another interesting feature disclosed by HST images is the presence of an  
unresolved spike at the center of the ''flat" cores of several (but not all!) 
giant E's (Forbes et al. 1995; Bower et al. 1997; Bower 1999). 

\subsection{Kinematics of E galaxies: global parameters}

An important review paper on the ''kinematical properties of 
early-type galaxies" 
is available (Capaccioli \& Longo 1994), giving precious details about the 
relations between the observable quantities and the actual kinematics of complex  
unresolved stellar systems, and also discussing the methods of data analysis. 
The emergent spectrum is the sum of the spectra of the stars along the 
line-of-sight: it may be 
approximated by the convolution of the spectrum of a star(s), so selected 
as to represent the dominant population (in terms of light output), by the 
line-of-sight velocity distribution (LOSVD). The problem is then to recover 
the LOSVD and to compare it with the one of suitable physical models of galaxy 
structure and dynamics. 

Representing the LOSVD by a Gaussian, one obtains its width, interpreted as a 
''velocity dispersion" $\sigma_{v}$, and its position in wavelength, which 
results 
from the line-of-sight velocity of the galaxy as a whole, plus local velocities, 
eventually consistent with a rotation $v_{rot}$ of the object. Recently, the
representation of the LOSVD by more than two parameters became feasible for
observations of improved S/N ratio: two Gauss-Hermite functions $H_3$ and $H_4$ 
of order 3 and 4 are added to the Gaussian (van der Marel \& Franx 1993; 
Bender et al. 1994). The term in $H_3$ describes an asymmetry of 
the LOSVD, expected for 
instance if two components with different velocities are superposed along the
line-of-sight. The term in $H_4$ sharpens or flattens the LOSVD against 
the Gaussian.  
Their comparison with the calculated parameters from dynamical 
models reinforce significantly the observational constraints on the theory of 
galaxy structures.  
 
Classical results may be summarized as follows:

a)  The central velocity dispersions of E's are rather well 
correlated with their luminosity: this is the Faber-Jackson relation which 
provided the first reliable estimator of the luminosities and distances of 
these objects (Faber \& Jackson 1976). 
It is now superseded by the Fundamental Plane (see Sect, 2.6), of which it is a 
projection. The central values of $\sigma_{v}$ are often used as substitute 
for the mass of E's.  

b) Again rotations are know for many E classified objects although much 
less than for the $\sigma_{v}$ parameter. 
When extending in a large enough range of $r$, the rotation velocities reach a 
plateau, and ''flat" circular velocities may be derived (Kronawitter et al. 
2000). 
Recent work dealing, with a statistically significant number of ellipticals,
and containing rotation as well as $\sigma_{v,0}$ data have been
published by Bender et al. (1994), Scodeggio et al. (1998) (aimed at 
the study of the FP), Mehlert et al. (2000) (Coma cluster). In a series of 
papers, 
Simien and Prugniel (1997 to 2002) extended kinematic data to faint E and dE  
(see also Halliday et al. 2001). 

Following Binney (1978), the kinematics of a galaxy is roughly described by the 
anisotropy parameter 
$(v_r/\sigma_m)^* = (v_r/\sigma_m)/\sqrt{\epsilon (1-\epsilon)}$. 
There $\sigma_m$ is the mean velocity dispersion within $r_e/2$ and $\epsilon$ 
the mean ellipticity. This gives the ratio $(v_r/\sigma_m)$ normalized to its 
value for an oblate spheroid of ellipticity $\epsilon$ supported by rotation.
The striking correlations of $(v_r/\sigma_m)^*$ with the morphological subclass, 
i.e. $di$E, $bo$E, $un$E, are described by Bender (1992) or in the review by 
de Zeeuw \& Franx (1991). 

The kinematics of the stellar population have been extended to large 
radial distances for a small number of objects and measurements outside
the $2r_e$ frontier have been made by Carollo \& Danziger (1994), 
Gerhard et al. (1998), Kronawitter et al. (2000) 
and Saglia et al. (2000a). 


\subsection{Surprising kinematical properties}

a)  Minor axis rotation
     
After the discovery in the seventies of the surprisingly slow rotation of 
several ellipticals, the eighties brought two developments who made ever 
more ''exciting" the kinematics, hence the dynamics, of such objects. The 
first was the detection of 
a large rotational velocity along the {\em minor axis} of NGC4261 
(Franx \& Illingworth 1988; Davies \& Birkinshaw 1988), 
instead of the usual major axis rotation of disks or oblate spheroids. 
Other cases of  kinematical misalignments were soon found 
and interpreted as a sure evidence of triaxiality of the systems. A 
discussion of the then available data is due to Franx et al. (1991). 
The misalignment angle $\psi$ is defined as $\tan \psi = v_{min}/v_{max}$, 
where $v_{min}$ and $v_{max}$ are the apparent velocities along the two axis. 
It is generally near zero, as for an 
oblate spheroid, but cases with large $\psi$ values are not infrequent,
exclusively among $bo$E or similar $un$E objects (see SAURON data in 
Copin et al., 2001).


b)  Kinematically decoupled cores

The second surprise was the finding of ''kinematically decoupled cores" (KDC) 
(Illingworth \& Franx 1988). 
These are small components, near the galaxy center, which do not share the 
apparent rotation of its main body. They are considered as resulting from 
the accretion of a minor outside galaxy.  
A remarkable example is NGC4365, a favored 
target to several observers. Davies et al. (2001) give results from the SAURON 
Integral Field Spectrograph, with 3D maps of the velocity dispersion, 
the apparent radial velocity $v_r$ and line strengths indices for the Mg$_b$ 
and H$\beta$ lines (see Fig. 6). 
The central 300 x 700 pc (or 3 x 7 arcsec) rotates about the projected 
minor axis, and the main body of the galaxy rotates almost at right angles to 
this. 
In this case the KDC has a disk-like geometry, both from the appearance of the 
$v_r$ map and from the fact that the innermost isophotes are disky, while they 
become boxy at about $r=5$ arcsec (see Carollo et al. 1997b). 
Such features are not infrequent and seem to occur both in $di$E and $bo$E 
objects (plus $un$E eventually). It is curious that no local anomaly of 
the stellar population are associated with the KDCs. This is indicated by 
V$-$I color measurements from high resolution HST frames  
(Carollo et al., 1997a). In the 3D maps of line strength indices and velocities 
from SAURON, the KDC and surroundings are found to 
have nearly identical stellar populations.

Integral-field spectrographs derived from the TIGER 
prototype of the CFHT (Bacon et al. 1995) are powerful new tools for the 
spectral analysis of extended sources, notably ellipticals. 
SAURON, for ''Spectroscopic Aerial Unit for Research on Optical Nebulae", has 
been attached to the William Herschel 4.2m telescope on La Palma, and has  
efficiently produced detailed 3D observations of the kinematics (plus other 
spectral properties) of a set of 72 galaxies selected from the start 
(Bacon et al. 2001, de Zeeuw et al. 2002). On the other hand OASIS, 
for ''Optically Imaging System for Imaging Spectrography", is associated 
with the adaptive optics system at the CFHT and may be therefore used for 
high resolution observations (Emsellem 1998; Bacon et al. 2000). 
A striking example of the potential of OASIS, plus the 
adaptive optics, is the discovery of a counter-rotating core of less than 2" 
diameter at the center of the $di$E galaxy NGC4621 in the Virgo cluster 
(Wernli et al. 2001). 


\subsection{Super Massive Black Holes} 

Super-massive black holes (SMBH) have long been invoked to explain the 
energetics 
of quasars and Active Galactic Nuclei (AGN). The quest of relic SMBH in local 
present 
day galaxies has been plagued however, with uncertainties associated 
with an insufficient 
angular resolution in the analysis of dynamical tracers close to the massive 
central 
object (see the review bye Kormendy \& Richstone (1995). 
In recent years, progress has been made on various crucial stages in the 
detection of 
SMBH in local E's, or in the bulges of S0 and a few spirals.
\begin{itemize}
\item High resolution imaging from the HST has been in systematic use, providing 
essential constraints on the geometry of the systems under study. The detection 
of minute gaseous rings around the center provided accurate Keplerian velocities 
about the central object, hence straightforward mass estimates (see review 
by Jaffe et al. 1998). For the famous active galaxy M87 (or NGC4486), the 
historical 
mass estimate of the central SMBH obtained from stellar kinematics 
by Sargent et al. (1978) has been rather nicely confirmed by the analysis of 
gaseous motions (Harms et al. 1994). 

\item Stellar kinematics at high angular resolution were also obtained recently 
with the HST spectrographs, notably the Space Telescope Imaging Spectrograph 
(STIS). 
Precise line of sight velocities distributions (LOSVD) were obtained close to 
the 
central objects, and the central peak in velocity dispersion accurately mapped. 

\item The LOSVD were analysed with elaborate dynamical models, such as the 
three-integrals axisymmetric models of van der Marel et al. (1998) or
Gebhardt et al. (2000b), although it has been 
suggested that the derived SMBH masses are not terribly dependent on the choice 
of the dynamical model. 
\end{itemize}   
   
The specialists are now confident enough in SMBH mass estimates to correlate 
these
with other galaxy properties. These correlation do not consider only 
ellipticals, but
also bulges of spirals notably the Milky Way and M 31. Quoting Magorrian et al. 
(1998) 
''a fraction of 97\% of early-type galaxies have MDOs (i.e. massive dark 
objects) 
whose masses are well described by a Gaussian distribution in 
$\log(M_{BH}/M_{bulge}$ 
of mean -2.28 and standard deviation 0.51". This is not a very tight 
correlation! 
More recently a better one has been found between SMBH masses and the galaxy 
velocity dispersion $\sigma_e$ (averaged within the half-light radius). 
Gebhardt et al. (2000a) give $M_{BH} = 1.2\pm.2 x (\sigma_e/200)^{3.75\pm.3}$, 
where 
$M_{BH}$ is in the unit of $10^8$ solar masses. The scatter of $M_{BH}$ is only 
0.3 dex (see also Ferrarese and Merritt 2000). 

Are the astrophysicists now entitled to discuss SMBHs, or should they continue 
to speak of MDOs, and remain prudent on the nature of the massive dark objects 
at the center of galaxies? When the central mass is encircled in a smaller and 
smaller radius, a larger and larger density is estimated, 
so that the only plausible physical nature 
of the MDO turns out to be the black hole. The study of E's does not afford 
opportunities to estimate in a compelling way the densities of their MDOs. 
A much better case is the Milky Way, where stellar kinematics constrain the 
density of the MDO to more than $10^{12}$ M$_\odot/\mathrm{Mpc}^3$,  
ruling out other physically acceptable objects than the SBMH.


Strong constraints are also found in the analysis of H$_2$O maser emitting rings 
near the nucleus of several Seyfert galaxies. 
Generally speaking however, the study of ''live" SMBH powering AGN (Active 
Galactic 
Nuclei) in Seyfert galaxies or quasars, relies on observable clues rather
different from those in the study 
of the ''dead" SMBH, and is outside the scope of this review. A beautiful 
summary 
of the relations of these domains of research is in Richstone et al. (1998). 
The following essential conclusions are quoted from this paper:
1) SMBHs are a normal feature of the central regions of bright galaxies, 
particularly those with spheroids. 2) Their masses scale in rough proportion to 
host-galaxy spheroid mass. 3) The total mass density in SMBH is broadly 
consistent
with the mass-equivalent energy density in the quasar light background. 
The association of SMBHs and spheroidal stellar systems suggests parent 
formation 
processes, probably in dense regions which collapsed early. On the other hand  
the ''broad consistency" in global energetics strongly suggests that present day
SMBHs are indeed relics of the powerful AGNs of the ''quasar era" around $z=2$ 
to 3.  
There remain discrepancies with the numbers however, the number of relics being 
too large compared to the number of parent quasars, possibly because statistics 
of 
ancient quasars select the brightest, while the statistics of the actual SMBH 
select those close to our terrestrial observatories. 
According to the quoted authors, one way out 
of these difficulties is to assume a progressive ''fattening" of black holes 
during 
and after the quasar era, so that many minor ancient AGNs were too faint to 
enter our quasar catalogues. 

The discovery of SMBH candidates in many spheroidal galaxies may further 
complicate
the problem of the origin and evolution of these objects. On the other hand it 
offers
a way to explain the nuclear profiles, ''core" against ''power-law", described 
above. van der Marel (1999) shows that these various profiles may result from 
the 
adiabatic growth of black holes into pre-existing isothermal cores, following 
a scenario first analyzed by Young (1980). This simple scenario is criticized by 
Merritt (1998) on theoretical grounds: he notes that the isothermal core is not 
favored by the theories of galaxy formation, either through hierarchical 
clustering 
or collapse. The end products of these processes are adequate nests to grow a 
black 
hole and develop a steep power-law nucleus (Quinlan et al. 1995, Merritt \& 
Quinlan 
1998). For the much less abrupt core-like profiles 
of giant ellipticals, the coalescence of two SMBHs following a merger of their 
host galaxies (Ebisuzaki et al. 1991, Makino 1997), is a preferred mechanism, 
notably 
for tenants of a real ''dichotomy" between the two kinds of central spheroidal 
profiles (Faber et al. 1997). 
 
\subsection{The Fundamental Plane}

Since the luminosity was known to be correlated with $SB_e$ and with 
$\sigma_0$  (Faber-Jackson relation), it is not surprising that a better 
correlation was found by using both observables simultaneously. 
The bivariate relation may be written as 
\begin{equation}
\mathrm{L} \propto \sigma_0^a(SB_e)^{-m} 
\end{equation}
or
\begin{equation}
r_e \propto \sigma_0^b(SB_e)^{-n} 
\end{equation} 
expressing the radius or luminosity in terms of observables. 
Similar relations have been published simultaneously by Dressler et al. (1987) 
and Djorgovski \& Davis (1987). Any one of these two relations 
define the so-called Fundamental Plane (FP) of elliptical galaxies. 

The FP theoretical basis is obtained from the virial theorem which gives 
M$=kRV^2/G$, where $R$, $V$ are suitable definitions of the  
radius and mean internal velocity, and G the constant of gravitation. 
Introducing observables and the M/L ratio, this becomes 
\begin{equation}
\mathrm{L.(M/L)}=kr_e\sigma_0^2
\end{equation} 
where $k$ is an ad hoc factor.
Assume now that M/L is a constant or a slowly varying function of the luminosity 
with $\mathrm{M/L} = L^{\beta}$. If E's
 are built along similar models 
the $k$ in eq. 3 is indeed a constant. Then using the definition of $SB_e$ 
one can obtain the relations (1) or (2). 
Note that the exponents in these equations 
can be expressed in terms of the slope $\beta$. 
As the M/L ratio varies with the colour, these exponents also vary.  

The value of the two-parameter relation, 
and the justification for the use of such a majestic vocable 
as Fundamental Plane, is the high quality of the fit: this might slightly be 
improved, by taking into account the role of rotation in the theoretical or 
empirical derivation of the FP (Prugniel \& Simien, 1994; Bender et al., 1994). 
As it is, the FP has many applications and 
physical interpretations. Thus it offers the possibility to 
estimate the distances of E's with 
an accuracy of about 15\%. For this particular application, it is 
often replaced by the $D_n$-$\sigma$ relation (Dressler et al. 1987), 
conceptually equivalent.

The quality of the FP relation implies that the M/L ratio of E's has 
only a smooth variation with L.
On the other hand, the structure parameter $k$ in eq. 3 should be 
nearly constant.
These questions have been studied first by consideration of the FP in the near 
IR 
(Pahre et al. 1995, 1998a) and later by comparisons of the FP from the U 
(0.35 $\mu$m) to the K (2.2 $\mu$m) spectral bands (Pahre et al. 1998b). 
It appears that metallicity variations are not sufficient to explain the 
changes of the FP slope with the colour. The problem has also been tackled by 
Prugniel and Simien (1996) who discuss the residuals from the 
FP against  
stellar populations indices (Mg$_2$ and colours). The effects of stellar 
populations explain only half of the apparent variations of M/L, i. e. the 
exponent $\beta$ above. The rest of the tilt of the FP should be explained by 
some dependence of the rotational support and of the spatial structure on the 
luminosity. The authors ''conclude to a constancy of the dynamical-to-stellar 
mass ratio". 

Various attempts have been made to find eventual differences in the FP for 
galaxies in different environments: poor clusters such as Virgo and 
rich clusters 
such as Coma; central core and outskirts of Coma (Lucey et al. 1991); field 
and cluster ellipticals (de Carvalho \& Djorgovsky 1992); cluster 
to cluster differences 
(J$\phi$rgensen et al. 1996); objects in {\em compact} groups against 
objects in the field or loose groups (de la Rosa et al. 2001). 
No salient variations in the parameters of the FP, or the dispersion about it, 
have been found. Prugniel et al. (1999) relate the residuals from the FP to a 
parameter $\rho$ describing the local density, and no convincing trend is 
present.    

\subsection{The baryonic content of E-galaxies}

\subsubsection{Star populations}

Since  stars cannot be resolved in any normal  giant elliptical, one has to 
rely on integrated colors and spectra to get information about their stellar 
populations. 
The classical Johnson's wide-band colors system has proved to be the most 
useful, 
eventually amended by the Cousins system. On the other hand,  integrated 
spectra are described by line-indices as, for instance, that defined by the Lick 
Observatory. A recent compilation of  Lick's data is given by  Trager et al. 
(1998).   
Colors and spectra of E's are similar to those of late GIII or early KIII 
stars.

There is a remarkable correlation between  colors and luminosities (or 
absolute magnitudes) of E's, brighter galaxies being redder. 
This so-called CM relation is best defined for 
clusters, and its slope is maximal for the color (U$-$V)  (Bower et al. 1992a, 
1992b). The CM diagram is considered to be a relation 
between 
the luminosity and the mean metallicity of the stellar population, the 
metallicity increasing with the luminosity. 
The existence of the CM relation and its small scatter  represent important 
constraints on the formation and evolution of E's. 

The explanation in terms of metallicity variations are also adequate for  
relations between  line-indices and  luminosity (or equivalently 
the $\sigma_d$ value). The ''best" relation occurs between Mg$_2$ 
(or Mg$_b$) and $\sigma_d$ also followed by bulges of spirals (Idiart et al. 
1996).  
There is also a fair correlation, but with an opposite 
slope, between the Balmer lines indices, such as H$_\beta$ and  luminosity, 
while the Fe-lines indices are nearly uncorrelated  (Worthey 1998).

Another important property of colors and line-indices in E's is their radial 
variation. Colors get bluer at larger $r$, metallicity indices become 
fainter, while  Balmer lines  become often stronger. About color gradients, 
one may quote the surveys of ''local" galaxies by Goudfrooij et al. (1994a), 
Michard (1999, 2000) and Idiart et al. (2002a). 

\subsubsection{Cold gas and dust}

It is an essential property of E-galaxies to contain very little dust or 
cold gas, compared to spirals, and accordingly no present day star formation. 
This is perhaps a reason why astrophysicists have been always very keen on 
detecting ISM in ellipticals.
Useful reviews are given by van Gorkom \& Schiminovich (1996) about the HI, 
Rupen (1996) for the CO, Goudfrooij (1996) about the dust and ionized gas. 

The survey by Goudfrooij et al. (1994a) deals with a complete sample 
(magnitude limited) of 
of 56 objects observed in BVI, and also in a filter isolating the 
H$\alpha+$[NII] 
emission lines. A comparison is then possible of the occurrence, localization 
and amount of dust and ionized gas (Goudfrooij et al. 1994b). Quoting the 
authors 
''The amounts of detectable dust and ionized gas is generally small, of order 
$10^4-10^5 \mathrm{M}_\odot$ of dust and $10^3-10^4 \mathrm{M}_\odot$ of 
ionized gas". These small amounts are detected in roughly 50\% of the 
ellipticals. 
Both gas and dust patterns tend to be centrally located, but without detailed 
correlation. 
 
Another survey based upon the B-R colours of 44 nearby E's has been 
published by Michard (1998a, 1998b, 1998c, 1999): it concentrates upon the
differences of dust content and distribution between $di$E and $bo$E, 
or between flat-core and power-law galaxies. 

The HST observations have brought important new knowledge of dust occurrence 
and geometry in the central region of E-S0 galaxies: Lauer et al.(1995), from 
V light maps; van Dokkum \& Franx (1995), from contrast enhanced images; 
Forbes et al. (1995) and Carollo et al. (1997), both from V-I maps. At improved 
resolution and S/N ratio, more and more minute dust features (lanes, filaments 
or dots) are detected, always near the center of the studied galaxies. 
Ellipticals with large organized dust systems, similar to those in later types, 
remain extremely rare, or atypical. 

It has been known since the publication of data from IRAS 
(Infrared Astronomical Satellite) ( Knapp et al. 1989) that dust is 
detected in many ellipticals from their thermal radiation at 60 and 100 $\mu$m.  
According to Goudfrooij \& de Jong (1995), far-IR estimated dust masses are 
often 
larger by an order of magnitude than dust masses calculated from the study of 
dust 
lanes and patches.  A diffuse dust component is postulated to bridge the gap.
The problem of diffuse dust in E's has been analyzed also by 
Witt et al. (1992) and Wise \& Silva (1996). A discussion by Michard (2000) 
suggests that its effect on color gradients is not significant.  
 
For normal early-type galaxies, 
the IR spectrum is in part due to the black-body radiation of cold stars 
(temperature near 3500 K); in part to hot dust (temperature around 300 K) 
presumably surrounding M giants, and to cold dust (temperature around 40 K) 
of doubtful localization. The Infrared Space Observatory or ISO, 
operational between November 1995 and April 1998, 
brought a wealth of new informations about all sources of IR 
radiation in the range 2.5-200 $\mu$m. 
Elliptical galaxies have been relatively neglected however, in the early work 
from 
ISO data. Preliminary results are given by Fich et al. (1998), 
Madden \& Vigroux (1999). The survey of the Virgo cluster by Boselli et al. 
(1998) 
contains only one giant elliptical.
Ferrari et al. (2002) are more generous with data for 28 early-type galaxies: 
mid-IR spectra, hints about the distribution, estimates of ''hot dust" masses. 
To sum up, while the existence of ''diffuse" dust in E's is certain, its spatial
distribution remains unknown. 

\subsubsection{X-rays and hot gas}

Strong X-rays emission was found with the Einstein X-rays Observatory 
from clusters of galaxies and from giant ellipticals (Forman et al. 1985; 
Canizares et al. 1987; Fabbiano et al. 1992). 
More powerful instruments, for both imagery and spectroscopy, were flown on 
ASCA (Awaki et al. 1994; Matsumoto et al. 1997), and on ROSAT 
(Davis \& White 1996). Further improvements were feasible for 
the instruments on board of 
{\em Chandra} and {\em XMM-Newton}, both launched in 1999, 
with an angular FWHM of a few arcsec for the later and 1 arcsec for 
{\em Chandra}.    
>From the recent review papers by Sarazin (1997) and Loewenstein (1999) we may 
summarize as follows the fascinating properties of giant E's in X-rays:
\begin{itemize}
\item The X-rays luminosities of E's (in the range $10^{39}-10^{42}$ erg/s) are 
correlated with their optical luminosities, but with considerable scatter 
(Canizares et al. 1987), and a logarithmic slope much larger than 1, i.e.
$L_X \propto L_B^2$.
\item The X-ray emission extends outwards in a large halo, with sizes of 
typically 
5-10 $r_e$. 
\item X-rays from E's are in part due to thermal emission from diffuse 
gas heated at some $10^7$ K or even more. 
Thermal equilibrium occurs between heating due to the 
large relative motions of the X-rays emitting stars, and radiative cooling. 
Incidentally, there is also a harder component due to bright stellar 
sources such as the X-rays binaries, which can now be individually detected with 
Chandra in Virgo cluster galaxies (Mushotzky et al. 2001). 
\item The emission lines in the X-rays range allow estimates of abundances 
of Fe, Si and a few other elements. The results can be reconciled with 
optical data (Loewenstein 1999). Abundances variations with redshift can now be 
studied (Mushotzky \& Loewenstein 1997)  
\end{itemize}

The hot X-rays emitting gas which fills many bright ellipticals is fatal to dust 
within a short time scale! Empirical consequences have been discussed by 
Goudfrooij (2000), such as the anti-correlation between the masses of dust 
and of X-rays bright gas.

\section{Cold Dark Matter Model}

\subsection{Evidences for the presence of Dark Matter}

For more than 20 years, the contribution of massive Dark Matter halos has been 
used to explain the flat rotation curves of spiral galaxies, both within 
the optical radius and in the extended HI envelopes (Bosma 1978;
Rubin et al. 1985; Sofue \& Rubin 2001). The relative contribution of the stars 
and DM to the total mass remains uncertain, specially towards the central 
regions. Improved kinematics (Palunas \& Williams 2000) may lead to ''no-halo" 
models within the optical radius, although the uncertainty in the M/L ratio of 
the 
stars leave room for some dark DM component.
The situation is not the same for dwarf galaxies, where a substantial amount of
dark matter is required (Hoekstra et al. 2001), with an average projected
density one order of magnitude greater than that of the neutral gas. In spite
of the uncertainties in the required amount of DM inside the optical radius, the 
rotation curves, either of spirals or dwarfs, indicate the presence of low 
density 
cores instead of the CDM cusps predicted by numerical simulations 
(de Blok \& Bosma 2002; Gnedin \& Zhao 2002; Marchesini et al. 2002).

The situation is somehow similar in the case of giant ellipticals: the variation 
of 
the M/L ratio is not sufficiently constrained, within the optical radius 
(typically 1 or 2 $r_e$), to ascertain eventual DM contributions. For instance, 
Ortega \& de Freitas Pacheco (1989) constructed self-consistent hydrodynamical 
models for sixteen galaxies, and concluded that most of them have an increasing 
M/L ratio with galactocentric distance, supporting the presence of DM. 
Binney et al.(1990), van der Marel et al. (1990) reached an opposite conclusion. 
Bertin et al. (1992) interpreted the kinematical data of 10 bright ellipticals
using collisionless self-consistent models and found some evidence for dark
matter. More recently, Ortega et al. (1998)
revisited the dynamics of a sample of ellipticals based on new kinematical data 
and 
on Jeans equations, allowing for oblate and prolate structures. Comparison of 
the resulting M/L ratios with
values derived from evolutionary models does not seem to leave room for an
important amount of DM inside the effective radius, not supporting their 
previous
conclusions. 

Such uncertainties have not been completely removed by the recent improvements 
in stellar kinematics (see Sect. 2.3). On the other hand, 
the number of E's surrounded by HI rings is quite small,  
and they are generally peculiar in some respect (van Gorkom \& Schiminovich, 
1996). 
For these objects, the determination of the amount of mass in a dark halo is 
similar
to the procedure adopted for spirals. Bertola et al. (1993) analysed five E's
combining the M/L ratio derived from dynamics of the inner ionized gas
and the outer neutral ring. They found a constant M/L ratio within
$r_e$, becoming very large in the ring region. 

Other kinematical tracers have been found useful:  

a) Planetary nebulae (PN) can be recognized in the 
outer galaxy regions of low surface brightness, and accurate radial velocities 
derived from the [OIII] emission of their envelopes (Arnaboldi \& Freeman 1996).
Arnaboldi et al. (1996, 1998), Mendez at al. (2001) have measured PN velocities 
in halos of several giant galaxies, giving further dynamical support to the
presence of a DM halo in those galaxies.

b) Globular clusters (GC) surrounds large galaxies in variable numbers and can 
be useful
tracers of the halo potential (Cohen and Ryzhov 1997; 
Kissler-Patig et al. 1998, 1999; Minniti et al. 1998).
 
In several cases, the information from two or more of the above techniques could 
be brought together to constrain dynamical models.
Several examples of increasing velocity dispersion at large galactocentric 
distances
have been found, implying an increasing  outwards M/L ratio due to a DM envelope 
(Kronawitter et al. 2000). A discussion of the 
data by Gerhard and al. (2001), indicates a variation of M/L with L in agreement 
with the ''tilt" of the Fundamental Plane.

An important method to study DM in ellipticals is based on the existence of
X-ray halos (see Section 2.7.2). The confinement of these hot X-ray coronae 
around
E galaxies requires a massive halo (Forman et al. 1985; Fabbiano 1989),
indicating M/L ratios in the range 10-80. Buote \& Canizares (1998) noticed
a different isophotal geometry for X-rays and optical light in NGC 720. The
X-ray isophotes are more elongated and their major axes are misaligned with
respect to the optical. If matter were distributed as in the optical light, it
could not produce the observed ellipticities of the X-ray isophotes and
they interpreted this effect as a consequence of the presence of a
massive DM halo. Also from
optical and X-ray data, Bahcall et al. (1995) concluded that the M/L ratio 
increases 
with radius up to values of about 100-150 M$_B(\odot)/\mathrm{L}_B(\odot)$ for
a sample of E's, these values being comparable with those found for clusters. 
More recent studies of microlensing
effects or Einstein arcs in clusters indicate mean $M/L$ ratios of about 350,
supporting a substantial amount of diffuse DM in clusters.

\subsection{Formation of large-scale structures and dark matter halos}

In the current scenario of structure formation, galaxies are assumed
to form, in a dissipative process, by cooling and condensation of gas, within 
dark matter halos (White \& Rees 1978; Blumenthal et al. 1984; White \& Frenk 
1991). 
The halos themselves form and evolve by purely gravitational mechanism.
Thus, on scales where the gas pressure is subdominant to gravity, 
the evolution of galaxies is governed by the evolution of the
underlying halos.  

It was initially believed that there exists two "contradictory" mechanisms for
the formation of structures in the Universe: top-down scenario 
consistent with a hot dark matter model and a hierarchical scenario 
in a cold dark matter (CDM) Universe. In the first scenario 
(Zel'dovich 1970)
flattened pancake-like structures 
form first on large-scales
(scales of order $10^{15} M_\odot$) 
and then fragment into
smaller clumps inside which galaxies are seeded. 
In the hierarchical scenario (Peebles 1965, 1972)
small objects would form first, nonlinearly interact
and merge to form larger halos which can then host galaxies. 
It is now understood that CDM model contains enough power on
large scales to be consistent with the first scenario as well as with the second 
one. Large-scale redshift surveys indicate that galaxies are distributed
on filamentary-type structures (Bharadwaj et al. 2000).
Numerical simulations
confirm that networks of giant walls (Zel'dovich pancakes) joined by
filaments, joined themselves at massive nodes, form on large scales at 
high redshifts in a CDM Universe. However, at much smaller scales the dynamics 
are nonlinear and virialized halos continuously form
by gravitational collapse and grow by accretion and merger.
 
On large scales, the evolution of dark matter
fluctuations is studied commonly using the equations of 
hydrodynamics for collisionless fluid.
To fix ideas, we consider an Einstein-de Sitter Universe
with $\Omega=1$. This provides a convenient context within
which we can discuss general concepts of structure formation.
In the comoving frame of an Einstein-de Sitter Universe, 
the continuity, Euler and Poisson equations are respectively written as,
\be
{\partial\delta\over \partial a}+{\bf\nabla}\cdot((1+\delta){\bf u})=0
\quad;\quad
{\partial{\bf u}\over \partial a}+{\bf u}\cdot{\bf \nabla}{\bf u}=
-{3\over 2a}\left({\bf \nabla}\varphi+{\bf u}\right)
\quad;\quad
\nabla^2\varphi={\delta\over a}
\label{euler}
\ee
where the density contrast is 
$\delta=(\rho-\bar\rho)/\bar\rho$ between the local density $\rho$ and the
background density $\bar\rho$, $a$ is the scale factor and ${\bf u}$ is
related to the peculiar velocity ${\bf v}$ by ${\bf u}={\bf v}/aH^2$
and $\varphi$ is related to the peculiar gravitational potential $\phi$ by
$\varphi=2\phi/(3 a^3 H^2)$.
In the linear regime ({\it i.e.}\ when $\delta<<1$) this system
of equations is solved analytically, yielding for the growing mode,
\be
\delta(t)={3\over 5}{a\over a_i}\,\delta_i,
\label{growth}
\ee
where  $\delta_i$ is the density contrast 
at the initial time $a_i$. Thus,
since the scale factor, $a$, grows as $t^{2/3}$, density contrast 
grows algebraically in the linear regime (on large scales). 

On scales where the dynamics are still linear, the gravitational and velocity
potential are equivalent; {\it i.e.} the right-hand-side of 
Euler equation in (4) vanishes. This means that the motion
of dark matter particles in this time frame (note that we have used the scale
factor $a$ as our time variable in eqs.(4) is basically
inertial. Zel'dovich used this result to demonstrate the formation of caustics 
(regions of infinite density bounding structures such as pancakes) in
self-gravitating systems; an element of dark matter fluid at an initial position
${\bf q}$ moves with its initial
velocity, ${\bf u}({\bf q})$, along straight line trajectories until it
crosses another element at which point a 
caustic forms. The continuity equation in (4) can also be written as
\be
\rho({\bf x},t)d{\bf x}=
\rho({\bf q}) d{\bf q}
\quad
;
\quad
\rho({\bf x},t)={\rho({\bf q})\over 
{\rm det}(d q_i/d x_j)}
\ee
where $\rho({\bf q})$ is the initial density at the initial particle position
and $\rho({\bf x})$ is
the density at the present particle position $({\bf x})$. 
The density field becomes infinite ({\it i.e.} caustics form)  when
the determinant in the above expression vanishes. The determinant can vanish due 
to
collapses along one, two or all three
axes results in the formation of pancakes, filaments and nodes respectively.
Numerical N-body simulations confirm the formation of such a network of
structures (see Fig.8) on large scales.
The formation of dark matter halos has been
related to singularities in catastrophe theory (Arnold 1986) 
and also to the maxima of the 
smallest eigenvalue (Shandarin \& Klypin 1984). 


In a cosmological context, bound dark matter halos
form within the aforementioned large-scale network where 
the density within a specific volume becomes 
higher than a critical density.
Therefore, typically galaxies are expected to form 
in the high density nodes, even though early-type
and faint dwarf galaxies can also form in overdense filaments. 

The criteria for the formation of dark matter halos is best understood within a 
spherical
model. Although far from spherical, typical dark matter halos
can be well-described by this simple approach. 
Place a sphere of radius $R$ randomly on fig. 8 at say redshift 10 
and assume 
a perturbation of 
uniform density, namely a top hat model. As the Universe expands,
because of its higher density the matter inside the sphere is decelerated
more than that outside, further increasing the density contrast. 
Eventually, it ceases expanding altogether and turns around to
recollapse. This event marks the transition to the nonlinear regime 
and occurs at a time (when $\delta=1$ in expression (5)
\be
t_{ita}={3\pi\over 4}\delta_i^{-3/2} t_i
\ee
If the collapse remains spherically symmetric, the central density becomes
infinite at a time $t_{c}=2t_{ita}$. Therefore, the linear density
contrast at the time of collapse is
\be
\delta_c={2\over 3\pi}\left({3\over 5}\right)^{3/2}\approx 1.69
\ee
Since in Einstein-de Sitter Universe
the density contrast grows as $(1+z)^{-1}$ (using $a\sim (1+z)^{-1}$ 
in  eq. (5)) one infers from the above expression that, for example, 
a region of density contrast $\delta_c=1.69/(1+z_{\rm dec})$ at the time 
of decoupling has just collapsed to from a bound halo.

The number of halos within a given mass range at a given redshift can be found
by using Press-Schechter theory (Press \& Schechter 1974).
In the CDM models the density fluctuation at decoupling is a
Gaussian-distributed field. Thus, within our randomly-placed spherical window, 
the probability that the density contrast exceeds the critical value $\delta_c$ 
is
\be
P_{>\delta_c}=
{(1+z)\over \sqrt{2\pi} \sigma}
\int_{\delta_c}^\infty
e^{-\delta^2 (1+z)^2/2\sigma^2} d\delta
\label{gaussian}
\ee
where the variance
\be
\sigma^2={\langle(\Delta M)^2\rangle\over\langle M\rangle^2}
=\int_0^\infty 
\left\vert\delta_k\right\vert^2W^2(kR) {d^3k\over (2\pi)^3}
\label{sigma}
\ee
is the rms mass fluctuation of the linear density field extrapolated to $z=0$
when filtered by the window function $W(kR)$ on a 
spatial scale $R$ containing mass $M$, and $\delta_k$ is the Fourier transform 
of
the density contrast. 
The results weakly depend on the window function which is often taken
to be a sharp k-space or a top hat filter.
Expression (9) gives the commulative probability of forming collapsed objects
with any mass. However, we are interested in the differential probability which
gives the number of halos within a particular mass range.
The number of halos, in the mass interval $dM$ is
\be
N(M,z) M dM=-2{\bar\rho}{dP_{>\delta_c}\over dM} dM
=
-{\sqrt 2\bar\rho\,\delta_c(1+z)\over\pi\sigma^2}
{d\sigma\over dM} 
\exp\left[{-\delta_c^2(1+z)^2\over
  2\sigma^2}\right] dM
\label{ops}
\ee
where the "fudge" factor $2$ 
multiplying the left-hand-side was originally
inserted by hand to ensure the right 
normalization for the total probability (Press \& Schechter 1974).
The halo mass function $N(M)$ depends on redshift through the variance $\sigma$.
For a scale-invariant power spectrum, $\vert\delta_k\vert^2\sim k^n$, the 
variance can be 
written as $\sigma=\left(M/M_*\right)^{-\alpha}$ where $\alpha=(n+3)/6$,
and the characteristic mass $M_*$ increases with decreasing redshift and is
defined by $\sigma(M_*)=\delta_c(1+z)$.
Hence, the mass spectrum of halos at any given time has the analytic form
\be
N(M,z)dM={1\over \sqrt{2\pi}}\left({n+3\over 6}\right)
{\bar\rho(1+z)\over M_*^2} \left({M\over M_*}\right)^{\alpha-2}
\exp\left[-(M/M_*)^{2\alpha}(1+z)^2\right]dM.
\label{mf}
\ee
The Press-Schechter mass function 
is universal: the shape of the distribution function is 
independent of the initial
mass function, redshift and cosmological parameters.
In spite of its simplicity and various unjustified assumptions, Press-Schechter
theory remains a reliable and influential analytic method for calculating the 
mass
function because of its remarkable
success when tested against numerical N-body simulations. 

Press and Schechter originally
developed their theory to describe galaxy formation by hierarchical
clustering. 
Subsequently, using a simplified version of the mass function, Schechter
(1976) constructed the fitting formula
\be
N(M)dM=\Phi(L) dL=\phi_*
\left(L\over L_*\right)^\beta
\exp\left(-L/L_*\right)dL/L_*
\label{lf}
\ee
for the galaxy luminosity function, where $\phi_*$ and $L_*$ are the
characteristic luminosity function and luminosity.
It is clear that mass and luminosity 
functions contain the same
information as long as one knows the value of $M/L$.
Thus, assuming a constant mass to light 
ratio for galaxies, one can directly compare 
the mass function (12)
to the observed luminosity function (13).
For small masses, the predicted slope of the 
mass function $(dN/dM\sim M^{-2})$ is a poor
fit to the power-law tail of the galaxy luminosity function, whose slope
depends on the galaxy colour selection and varies between $dN/dL\sim L^{-3/2}$
in the blue and $dN/dL\sim L^{-1}$ in the red end of the spectrum.
Thus, hierarchical models over-predicts significantly the number of
satellite galaxies ({\it e.g.} $\%95$ of the satellites of milky way
predicted by this model have not been observed).

In arriving at the Press-Schechter mass function, we have
overlooked the fact that a halo can contain smaller sub-halos, 
which is referred to as the cloud-in-cloud (CIC) problem. 
The CIC problem has been dealt with
in an extension of Press-Schechter theory (Bond et al. 1991).
In the extended theory the factor of 
$2$ in (\ref{ops}) is derived naturally and the agreement 
between the predicted and observed
luminosity functions at the faint end of the spectrum is improved,
since the number of low-mass objects 
diminishes due to merger (White and Frenk 1991).
Recently, a different fitting function  
has been obtained directly from numerical simulation results 
(Sheth and Tormen 1999). The discrepancy between theory and simulation
can be due to the assumption that 
objects collapse spherically in the Press-Schechter theory. Indeed,  
it has been shown that this discrepancy reduces substantially if bound 
structures are assumed to form from an 
ellipsoidal, rather than a spherical, collapse (Sheth et al. 2001).

\subsection{Density profile of dark matter halos}

It is generally believed that galaxy rotation 
curves depends on the density profile of dark matter halos.
Analytic evaluation of the halo
density profile goes back to the work of 
Gott (1975) and Gunn (1977) who established that
gravitational collapse could lead to the formation of virialized systems with
almost isothermal density profile ($\rho\approx r^{-2.25}$). 
Their work was based on the assumption that halos form by accretion or 
secondary infall of matter onto an initially overdense perturbation. If infall
produces density profile as shallow as $r^{-2}$, then the flat rotation curves
of disk galaxies and the related observation of increasing $M/L$
on large scales might thereby be explained. A somewhat steeper
$r^{-3}$ profile resulting from infall also explain the Hubble law light
profile of elliptical galaxies (Gott 1975).
Further refinement was made (Bertschinger 1985) and it was shown analytically 
that the
secondary infall approaches a self-similar form, whose exact behavior depends on 
the
central boundary conditions and on the kind of gas; if there is no
central black hole, the form proposed by Gott \& Gunn prevails, otherwise
the density profile has the power-law form $\rho\approx r^{-1.5}$.

That the halo profile should vary from one cosmological model to another
was proposed later on (Hoffman \& Shaham 1985).
Approximately flat circular velocities were then obtained for $\Omega\sim 1$ and
$-2 < n < -1$ (where $n$ is the spectral index, see section above expression
(12). This provided a strong support for the then
fashionable CDM model where at the galactic scale the index of
the power-spectrum was  $-1.5$ (Blumenthal et al. 1984).
It therefore became customary to model virialized halos by isothermal spheres
characterized by two parameters; a velocity dispersion and a core radius.
Later on, in a high resolution simulation of a galaxy-sized CDM halo, 
surprisingly, it was found that the density profile 
could be singular even at very small scales 
(Dubinski \& Carlberg 1991) and well-approximated by the model proposed
for the elliptical galaxies (Hernquist 1990) 
for which $\rho\sim r^{-1}$ in the inner parts, down to the
smallest resolved scale, $r\approx 1$ kpc. 
Further on, it was shown 
(Navarro et al. 1996, 1997) that
the spherically-averaged equilibrium density profiles 
of CDM halos of all masses can
be fitted by the formula
\be
\rho(r)={\rho_s \over (r/r_s)^{\alpha}(1+r/r_s)^{3-\alpha}}
\label{nfw}
\ee
where is this case $\alpha=1$, $r_s$ is a scale radius where the profile changes 
shape 
and $\rho_s$ is a characteristic density.  
Although, this profile was originally found for a standard CDM
model, it turned out that (14) provides an excellent fit to halo
density profile from other cosmological 
models (Cole \& Lacey 1996). Expression (14)
is now assumed to be a universal two-parameter function describing the profile
of dark matter halos which is independent of the halo mass, redshift, the
initial density fluctuation spectrum and the values of the cosmological
parameters.

Therefore, at small radii, the results of numerical simulations
indicate that the density profile (14) might be cuspy 
without the presence of a central core. Although an isothermal profile 
at large radii agrees with the galaxy rotation curves, there are 
controversies as whether a cuspy profile near
the centre is in accordance with various observations such as the rotation-curve 
of
well-studied gas-rich dwarf spirals (Flores \& Primack 1994). In addition, it 
has been 
argued that the radial arcs, require a
flat core in the cluster density profile, in apparent contradiction with a
singular density profile (Mellier et al. 1993).
Subsequent independent higher-resolution simulations (Fukushige \&
Makino 1997; Moore et al. 1998)
have confirmed the presence of a central cusp, but a shallower
density profile ($\alpha=1.5$) has been obtained at small radii 
which might resolve some of the aforementioned problems.


\section{Galaxy formation}

Galaxy formation is a complex process in which not only gravity, discussed in
the previous sections, but
also gas dynamics, chemistry and turbulence play crucial r\^oles. 
We still do not have a clear understanding of galaxy formation
and the reason for this predicament is that we do not yet have a fundamental
theory of star formation. Most theoretical studies in this field are based on
phenomenological arguments and numerical simulations.
Among the different processes involved in galaxy formation are those
responsible for heat balance of the gas: shock heating and radiative
cooling within collapsing dark matter halos, the subsequent
transformation of cold gas into stars and the regulation of star formation by
feedback from stellar winds and supernovae. 
Application 
of CDM theory at the galaxy scale is rather difficult and most results 
are based on numerical studies which are mainly either direct
simulations or semi-analytic modelings of galaxy formation. 
In the first approach the gravitational and hydrodynamical equations are
solved explicitly using various types of numerical codes
({\it e.g.}\ Katz et al. 1992). In the second approach, the
evolution of baryonic gas is calculated using simple analytic models while the 
evolution
of dark matter is evaluated using numerical N-body simulations (White \& Frenk
1991) or a Monte Carlo techniques in which one studies the merger tree of 
dark matter halos (Cole et al. 1994) .

\subsection{Heating \& cooling of gas}

During the collapse of a dark matter halo, the baryonic gas is supposed to
be shock-heated,
or undergo adiabatic compression depending on the halo mass, to the
virial temperature $T_{\rm gas}$ of the host halo, given by
\be
T_{\rm gas}={2\over 5}{GM\mu m_{\rm h}\over k R_{\rm vir}}
\ee 
where $M$ is the total mass contained inside the virial radius
$R_{\rm vir}$ which is usually taken to be half of the initial turn-around
radius, $\mu$ is the molecular 
mass and $m_{\rm H}$ is the mass of hydrogen atom.
Typically
$T_{\rm gas}=7.17\times 10^{6} \mu M_{12}/R_{30}$, 
where $M_{12}=10^{12}M_{\odot}$ and $R_{30}=30 kpc$.

To form stars the gas has to cool. The characteristic cooling time, defined
as the ratio of the thermal energy density to the cooling rate per unit volume 
$\rho^2_{\rm gas} \Lambda(T_{\rm gas},Z_{\rm gas})$, is
\be
{\tau}_{\rm cool}=
{3\over 2\mu m_{\rm H}}{kT_{\rm gas}\over 
\rho_{\rm gas} \Lambda(T_{\rm gas}, Z_{\rm gas})}
\ee
where $\rho_{\rm gas}$ is the density of the gas, $Z_{\rm gas}$ is the
metallicity and $\Lambda(T_{\rm gas}, Z_{\rm gas})$ is the cooling function, 
normally taken from tabulated values 
({\it e.g.}\ Sutherland \& Dopita 1993).
A second constraint on star formation is imposed by the dynamical time
, $t_{\rm dyn}=(\pi^3/6 G\rho)^{1/2}$ 
which has to be longer than the characteristic cooling time. If we
suppose that the gas has a primordial chemical 
composition, {\it i.e.} $\mu\sim 0.59$,  then the
cooling process is essentially due to bremsstrahlung and the cooling
function is proportional to $T^{1/2}$. In this case, equating both
timescales one obtains a critical density 
$\rho_{\rm crit}\ge 1.6\times 10^{-13} g cm^{-3}$ which for the above example
 gives a temperature of $T\sim 4.2\times 10^{6}$ or 
$\rho_{\rm crit}\ge 6.7\times 10^{-26}$ g cm $^{-3}$. 

If we assume, in a first approximation, that baryon distribution follows the 
dark matter distribution,
then $R_{\rm b}\sim R_{\rm DM}$ which gives
a critical baryon mass of
\be
M_{\rm crit}^{\rm b}={4\pi\over 3}\rho_{\rm crit} R_{\rm DM}^3 \sim 6\times 10^9 
M_\odot
\ee
corresponding to a dynamical time of 
$t_{\rm dyn}\sim 0.034/H$ where $H$ is the Hubble constant. Thus, only
fragments heavier than the above mass can cool fast enough 
to fragment and form stars during gravitational collapse.

Star formation converts gas into luminous matter and at the same time affects
the physical state of surrounding gas, since supernova explosion and young
stars inject energy and metals out of the galactic disc in the form of a hot
wind. The latter acts as a feedback process which regulates the star formation
rate. In addition, the injected metals enrichen both the cold star-forming gas
and the surrounding diffuse hot halo gas.
Thus, the later evolution of galaxies depends strongly on when the first stars 
form because
they can reionize the intergalactic 
medium (at $z\approx 7$) and enrichen it with metals, affecting the
cooling-heating balance. Metals increase the cooling efficiency, decreasing
the gas cooling time and increasing the star formation rate.  
In the hierarchical approach, small objects form first and, if constituted
only of primordial material, the cooling timescale may be longer than the
Hubble timescale, being one of the problems in this scenario.

\subsection{Angular momentum and morphology}

Apart from the clear difference in their stellar populations, the other
difference between spiral disks and ellipsoids is that the former are supported
against gravity by rotation, whereas for the latter it is primarily the {\it
pressure} of the anisotropic stellar orbits that is responsible.
The significance of rotation
may be quantified by a dimensionless {\it spin parameter}
which is the ratio between the observed angular velocity
 and the angular velocity which would be required
to support the galaxy by rotation alone:
\be
\lambda={J\vert E\vert^{1/2}\over G M^{5/2}}
\ee
where $E$ is the binding energy and $J$ is the angular momentum.  The spin
parameter can be as high as $\lambda\sim 0.4$ for spirals and only $\lambda\sim
0.05$ for ellipticals.

A physical explanation of these values remains a major unresolved
problem.   
Recent N-body simulations show that in the absence of star
formation, protogalactic gas loses much of its angular momentum as it cools
into pregalactic fragments that subsequently merge, producing disks that
rotate much too slowly 
(Navarro et al. 1995; Navarro \& Steinmetz 1997) which is a 
difficulty concerning the possible formation of S0's and E's from the
merger of spirals.

A possible but poorly-understood solution is feedback, briefly
discussed previously, which can prevent the gas from cooling into dense
fragments.
Furthermore, the Kelvin circulation theorem guarantees that in the absence of
dissipation, an initially irrotational velocity field remains irrotational. 
Earlier on we
have seen that on large scales motions of self-gravitating particles are
in the form of a potential-flow which has zero vorticity. 

A possible origin for the rotation of galaxies was suggested,
for the first time, by Hoyle (Hoyle 1949). In his
model the acquisition of angular momentum is attributed to the tidal action of
protogalactic
objects around it, at the epoch when the protogalaxy is
just about to form a galaxy. One assumes that the initial density field 
consists of non-spherical lumps with mass $M$ 
and density contrast $\delta$. The total torque, ${\cal T}$, produced on a
clumps, which is roughly taken as a dipole, from another one at a distance of 
$R$ is
$
{\cal T}\sim 
{\delta G M^2/ 4 R a}
$
where as before $a$ is the scale factor. In the linear regime $\delta\sim a$
 and hence the torque remains constant and the angular momentum
grows linearly with time ($J\sim t$). To obtain the binding energy
once again we invoke our simple spherical model. We know
the density contrast at the time of collapse is ${\bar\rho/\rho}\sim 5.6$.
Since the mass of the object is $M\sim (R/2)^3 4\pi \rho_0/3$, this
gives the proper radius of the object at turn around in terms of which the
binding
energy can be written $E=-3GM^2/5 r_{\rm prop}$. Let us assume that the
  object has
an overdensity $\delta_c$  then we obtain for the dimensionless spin
parameter
\be
\lambda\sim {\delta_c\over 40}\sim 0.04
\ee
which is in very good agreement with that of ellipticals but is too low to
account for the angular momentum spiral galaxies. But how can one account 
such morphological segregation ?
 A possible explanation is that ellipticals formed earlier when the universe
 was
denser and angular momentum smaller. An alternative explanation
frequently used in CDM hierarchical scenario, is that spirals formed first and
ellipticals formed from mergings of the spirals. This might seem to be
plausible given that ellipticals are usually observed in the dense regions,
but as we shall see later, there are severe
observational evidences disfavoring this picture.

It is worth commenting that dissipative effects, neglected in the tidal
picture, might be important factors in the explanation of the origin of the
angular momentum and its morphological dependence. However, dissipative
mechanisms are complex processes and so far rather poorly understood.
Although, angular momentum is approximately conserved during baryon infall
and disk accretion, the numerical simulations (with CDM model) show that most
of the angular momentum is actually lost to the dark matter halo. The
clumpiness (existence of substructure) induces strong angular momentum
transfer via
tidal torquing and dynamical friction from the dissipating baryons to
energy-conserving collisionless dark matter. The resulting disk therefore has
very little specific angular momentum (Navarro \& Steinmetz 2000). A possible
solution is feedback by supernova explosion which would
inject angular momentum into infalling baryons, which however,
leads to significant delay in the formation of disks.

\subsection{Galaxy mergers}

We have already discussed that during their evolution dark matter halos merge
and form larger halos with masses which can host the galaxies. However, there
is a second important merger event and that is the merger of the galaxies 
themselves. 
In hierarchical models of galaxy formation, it is normally assumed that when 
dark 
matter halos merge, the most massive galaxy automatically becomes the central
galaxy in the new halo while the other galaxies become
satellite galaxies orbiting within dark matter halo.
The orbits of these satellite galaxies will gradually decay as energy and
angular momentum are lost via dynamical friction to halo material. Thus
eventually the satellite galaxies will 
spiral in and merge with the central galaxy.
 
In these models, the galaxy morphologies are
ascribed to the particular merger history and accretion events that galaxies
experience during their assembly in the Universe.
In this model, disks are envisioned to form as the result of gas accreted
smoothly from the intergalactic medium (Katz \& Gunn 1991) whereas the
primary route by 
which elliptical galaxies and the bulge
components of spiral galaxies form is through major merger events where disks
are thrown together and mixed violently on a short timescale (Toomre 1977).
Some numerical N-body simulations have shown that merger of galaxies of
comparable masses (major mergers) result in the formation of elliptical
galaxies (Barnes 1988).

We comment that one can have a monolithic scenario of 
galaxy formation which accommodates the merger
of halos but not that of the galaxies. For this, collisions
between halos must not be head-on and the dynamical friction timescale
between baryonic and dark matter should be larger than   
Hubble time. 
  
\subsection{The epoch of galaxy formation}

The epoch of formation of a galaxy is usually defined as the redshift 
by when more than half of its present mass had been assembled within a 
sphere of radius  $\sim 30$ kpc.

As we have mentioned earlier, the main competing beliefs 
for the epoch of galaxy formation are monolithic
(high-redshift formation time) and hierarchical (low-redshift formation time). 
The simplest possibility for star formation epoch is that the baryonic gas 
turned into star at the very early stage of the collapse of dark matter.
There would then be no segregation between stars and dark matter, since stars
would behave as collisionless particles. It was then argued that
metal-poor stars in the solar neighborhood have very eccentric
orbits and hence must have formed in a time short compared to the collapse
time, such that their orbits mixed collisionlessly 
(Eggen et al. 1962). This is one simple description of
the monolithic scenario.

Application 
of CDM theory at the galaxy scale is rather difficult and most results 
are based on numerical studies which are mainly either direct
simulations or semi-analytic modelings of galaxy formation. 
In the first approach the gravitational and hydrodynamical equations are
solved explicitly using various types of numerical codes
({\it e.g.}\ Katz et al. 1992). In the second approach, the
evolution of baryonic gas is calculated using simple analytic models while the 
evolution
of dark matter is evaluated using numerical N-body simulations (White \& Frenk
1991) or a Monte Carlo techniques in which one studies the merger tree of 
dark matter halos (Cole et al. 1994) . 

A tight upper-bound on the formation redshift of galaxies can be obtained by 
using 
again our simple spherical model for a protogalaxy of 
radius $r$.  A protogalaxy can only be formed 
when its mean average density, $\bar\rho$, is higher than the mean background
density $\rho_b$. Dynamical relaxation after formation 
increases the internal energy of the stars in the protogalaxy. Thus a reasonable 
bound on 
the circular velocity, $v_c$, (in the case of a disk system) or on the
dispersion velocity, $\sigma_d$, (in the case of a ellipsoidal system) is the 
present value for the bright galaxies. 
Using $H=(8\pi G\rho_b/3)^{1/2}\sim H_0\Omega_m^{1/2}(1+z)^{3/2}$, we get the 
ratio
\be
{\bar\rho\over \rho_b}={2 v_c^2 \over \Omega_m(H_0 r)^2(1+z_f)^3}
\ee
which has to be larger than unity at the formation time.
For a spherical model this ratio can be estimated easily.
The proper radius $r(t)$ of the spherical shell containing 
mass $M$ has the parametric form (Peebles 1980)
\be
r(t)=A(1-{\rm cos}\theta)\quad;\quad
t=B(\theta-{\rm sin}\theta)
\ee
where $A^3=GMB^2$.
The spherical shell initially expands with the background while decelerating
more due to its higher density, until it reaches the maximum 
radius at $\theta=\pi$ and turns around
and collapses. In Einstein de-Sitter Universe the background density
falls as $1/6\pi G t^2$, and hence at the time of maximum expansion the ratio
\be
{{\bar\rho}\over\rho_b}={9\pi^2\over 16}=5.6
\ee
is obtained. 
Equating this ratio with (20) and 
for a circular (dispersion) velocity of $v_c(\sigma_d)\sim 200 {\rm kms}^{-1}$ 
and $r\sim 30 h^{-1}$ kpc we obtain
\be
z\le 10\Omega_m ^{-1/3}.
\ee
Thus, for $\Omega_m\sim 0.3$ Universe, bright spheroids of galaxies 
could be in place at redshift of about $z\sim 10$.

Thus, arguments such as above admit the possibility that galaxy formation could 
occur
relatively early at redshifts up to $10$. However, in CDM models galaxies are 
expected to form at redshifts much lower than this. One reason for this is
that the clustering pattern of galaxies, as measured by two-point correlation
evolves rapidly with time in Einstein-de Sitter Universe. If galaxies were to
form at $z=10$ then one would expect a drastic steepening of the slope of the
correlation function which is incompatible with observations.

\section{Stellar populations and chemical evolution}

\subsection{Stellar population synthesis and applications}

An important point should be emphasized: the interpretation of 
the integrated colors or line indices of complex stellar populations, 
requires necessarily the use of models. Two main "schools" have been 
practicing this exercise in the past years: the first use single stellar 
population 
(SSP) models, the second evolutionary models.

 SSP models are straightforward applications of the theory of stellar evolution. 
Given an initial mass function (IMF) and chemical composition, 
the theory allows to calculate the time evolution of such a population and the 
corresponding spectral properties are obtained from the available stellar 
''library" (theoretical or 
empirical). The uncertainties in such models were discussed by Charlot et al. 
(1996). 
The comparison with the observed indices (color or lines) give 
constraints on the input parameters (age and metallicity in the
simplest case). A popular series of SSP models by Worthey (1994) has been 
extended 
by Trager et al. (2000b) to take into account non-solar ratios of 
$\alpha-$elements 
to Fe, a bias in usual spectral libraries first noted by Borges et al. (1995).   

Evolutionary models are, in general, of the ''one-layer'' type including (or 
not)
mass loss from galactic winds, or infall of matter from the intergalactic 
medium. 
The star formation begins when a critical gas mass is attained (see Section 4), 
corresponding to a galactic age of about 15 Gyr. 
The population mix is calculated, as a function of time, by summing the SSP 
models representative of successive stellar generations: 
these are controlled by the amount of residual gas available for star formation, 
while their chemical composition results from the progressive enrichment of the 
ISM 
in ''metals" by gas loss from stars, notably supernovae. Evolutionary models 
have been
calculated by Bressan et al. (1996); de Freitas Pacheco (1996a); 
Vazdekis et al. (1996); Kodama \& Arimoto (1997). The model used by  
Idiart et al. (2002b) is an upgrade of the quoted one by de Freitas Pacheco 
(1996a).    
To obtain representative populations for E's of various masses, one selects 
input parameters, i.e., the star formation efficiency and the IMF, through 
an iterative procedure, in such a way as to reproduce one fundamental property 
of E's (or eventually several). 
For instance, in the quoted models by Idiart et al. (2000b)  models 
are fitted, through an iterative procedure, to the color-magnitude diagram 
(U-V) versus $M_V$ of ellipticals in the Coma and Virgo clusters. 
The required star formation efficiency should then be 
an increasing function of the galactic mass, and the slope of the IMF 
should be slightly flatter in brighter objects.

With the advent of fast computers and powerful simulation techniques, there has 
been
considerable progress in describing the growth of structures and in reproducing
some of the basic properties of galaxies and their evolution. In the case of 
E's,
codes including Smooth Particle Hydrodynamics and N-body Tree, have been used to
explore whether the formation of E-galaxies inside non-rotating virialized DM 
halos
leads to results compatible with actual data. Models along these lines, 
including 
heating and cooling
(see Section 4.1), energy feedback and chemical evolution have been developed
(Chiosi \& Carraro 2002), indicating that they can reproduce the basic observed
properties of E-galaxies as the color-magnitude relation or the $M/L$ ratio 
versus
central velocity dispersion. These simulations suggest that
massive E's form the bulk of the stars in a short timescale, whereas low-mass 
E's
have a complex star formation history: in other words, star formation is early 
and "monolithic" in high initial density systems, irregular and intermittent in 
low initial density ones. With the models by Kawata (2001), 
the low $v_{rot}/\sigma_v$ ratio observed in bright ellipticals
as well as their brightness profile ( the $r^{1/4}$ law), can be reproduced 
adequately by simulations considering the clustering of subclumps caused by 
initial small-scale density fluctuations. This phenomenon leads to angular
momentum transfer from the baryon component to the DM, resulting in a nearly
spherical system supported by random motions. On the other hand, N-body/SPH 
simulations eventually predict that the morphology (see Section 4.2) 
is a "transient" phenomenon within
the lifetime of a galaxy and is determined basically by the mode of gas 
accretion
(Steinmetz \& Navarro 2002). Disks arise from the smooth deposition of cold gas 
at
the center of DM halos while spheroids result from the stirring of preexisting
disks during mergers. 

\subsection{Inferences from models}

Age and metallicity effects can be disentangled using SSP models 
if different indices (colors or/and spectral) are used 
and if the considered stellar system is {\em really single} 
(de Freitas Pacheco 1997, 1998). Elliptical galaxies are 
certainly not single systems. If, on the one hand the age range of the 
population mix 
constituting the galaxy is probably quite narrow (3-4 Gyr), a reason 
invoked to justify the use of SSP models, on the other
hand the build up of chemical elements requires successive stellar generations,  
producing a  metallicity spread adequately described by population mix models.

Nevertheless, SSP models have been successfully used to get insight 
in the chemistry of ellipticals. A summary of analysis made on different mean 
elemental abundances in E's can be found in Worthey (1998). 

In the last decade many studies suggested that
the mean [Mg/Fe] ratio in bright ellipticals, derived from the comparison 
between observed and synthetic indices, are non-solar (Worthey et al. 1992; 
Carollo \& 
Danziger 1994; de Freitas Pacheco 1996a). The resulting enhanced [Mg/Fe] ratio 
could simply be the consequence of a short period of star formation activity, 
where the bulk of the stellar population is formed, with successive 
stellar generations being enriched mainly by the ejecta of type II supernovae 
(SNII). 
The iron-rich ejecta of SNIa begin to contaminate the medium later, 
about 0.8-1.0 Gyr after the onset of the star formation process, and only a 
small fraction of the total stellar population can  be enriched by these events, 
producing stars with lower [Mg/Fe] ratios, i.e. near solar. 

Using SSP models, Trager et al. (2000b) derived from the indices  H$\beta$, 
Mg$_b$ and $<Fe>$ an average (near center) metallicity [Z/H]$=+0.26$, 
for a sample of 39 E's. Kobayashi \& Arimoto (1999) used the same models 
(but with solar [Mg/Fe]) and the indices  $<Fe>$ and Mg$_2$ (Mg$_b$) to obtain 
mean metallicities for a sample of 80 early-type galaxies. 
The resulting values are in the range $-0.8 < \mathrm{[Fe/H]} < +0.3$, with the 
abundances derived from the  $<Fe>$ index being systematically  lower 
than those estimated from the Mg$_2$ (Mg$_b$) index. 

Idiart et al. (2002a) obtained central colors (UBVRI) and gradients 
for a sample of 40 early-type galaxies, and derived central and mean 
metallicities using these data and evolutionary models described above 
(Idiart et al 2002b). They find  mean metallicities
in the range $-0.21 < \mathrm{[Fe/H]} < +0.24$, with a sample 
average [Fe/H]$ = +0.01 \pm 0.11$(rmsd). Central values are about $0.28$ 
dex higher with nearly the same dispersion. These
results suggest that ellipticals have a narrow range of metallicities, 
varying within a factor of three, while their masses vary within two orders of 
magnitude (from $2 \times 10^{10}M_{\odot}$ to $2 \times 10^{12} M_{\odot}$), 
according to the models of the aforementioned authors.

While both SSP and evolutionary models tend to agree in the determination of the 
[Fe/H] and [$\alpha$/Fe] ratios, there are still disagreements in age 
determinations.  Trager et al. (2000a) 
determine ''SSP-equivalent ages", together with metallicities and [$\alpha$/Fe] 
enhancements from three line indices, and accept  real age 
excursions from about 2 to 24 Gyr.    

Colours and spectral indices from evolutionary models 
of ellipticals, including gas inflow and outflow were computed by 
Bressan et al. (1996), who concluded that E-galaxies do not follow a 
pure sequence either of age or metallicity: 
galaxies of different masses have suffered different star 
formation histories, in spite of having rather old (13-15 Gyr) 
populations. Kodama \& Arimoto (1997)
considered two model sequences (age and metallicity sequences), in which the
parameters were adjusted in order to fit the CM diagram of ellipticals in 
Coma. They compared the evolution of both CM sequences up to redshifts $z \sim 
1$ 
and concluded that the age sequence deviates significantly from observations 
already at $z \sim 0.3$, supporting the view that the CM diagram is primarily a 
metallicity variation with luminosity. Similar conclusions were reached by 
Tamura et al. (2000) from the analysis of E's in the 
Hubble Deep Field North, as well as by Saglia et al. (2000b), who investigated 
the evolution of colour gradients with the redshift, by comparing values in the 
cluster CL0949+44 ($z \sim 0.37$) with those derived for local ellipticals.  

In the series of evolutionary models published by Idiart et al (2002b), the star 
formation 
activity begins about 15 Gyr ago and the mean ages of the population mix vary 
from 12.5
to 13 Gyr in the full range of galaxy masses.  In these old populations, the 
bulk of the stars 
were formed within a time  interval of about 2 Gyr, or at an epoch 
$z \sim$ 2.5 -3.0 (for a world model with $\Omega_{m}$ = 0.3 ,
$\Omega_v$ = 0.7  and  $H_0$ = 65 km/s/Mpc). The calculated spectral properties, 
notably the relations between the H$\beta$ and other line or color indices, are 
in general agreement with observed data.    More recently, Waddington et al. 
(2002)
obtained images with the HST of  two previously known elliptical galaxies 
(53W069
and 53W091) at z $\sim$ 1.5, having old stellar populations. Their observations
indicate that the rest frame (U-V) colour gradient of 53W069 is consistent with 
that of
present-day E's, suggesting that the bulk of the stars were formed on a very 
short
timescale at z $>$ 5. The other object has a colour gradient larger than those
observed in ellipticals at z $<$ 1, but the authors concluded that also in that 
E-galaxy
the bulk of the stellar population was formed in a high-redshift short star 
formation burst.
Waddington et al. concluded that both galaxies are  ''passively'' evolving (see 
Section 5.3)
into ordinary ellipticals by the present day.

\subsection{Cosmic evolution}

In this section we review  the main sources of information presently available 
to guide our efforts towards a better understanding of the formation and 
evolution of 
E-galaxies.

In the hierarchical scenario, massive E's form at relatively low
redshifts ($z \leq 1.5$) through the merging of spiral 
galaxies, so that older ellipticals should be rare at higher redshifts. In 
contrast, the monolithic scenario predicts that massive ellipticals 
formed at $z  > 3$, after a brief and intense star formation period and 
are evolving passively, implying a constant comoving number density. 
In this case, a substantial number of objects with the integrated colours and 
spectra of old populations is expected to be detected at $z \approx 1$. 
There has been however a number 
of discrepant results: some authors found a clear deficit of old objects at $z 
\sim 1$ 
(Stiavelli \& Treu 2000; Rodighiero et al. 2001), while others found evidence 
for a constant comoving density (Scodeggio \& Silva 2000; Im et al. 2001).
>From near IR surveys, Extremely Red Objects (EROs) have been known for years, 
and 
their identification with either star forming dusty galaxies or old early-type  
galaxies with passive evolution has been proposed. Recent work, using relatively 
large survey fields, aimed at choosing between the two possibilities using 
the R$-$K colors, spectra and clustering properties. It appears that both
proposed types of EROs occur in roughly equal proportions at $0.7 < z < 1.5$ 
with an age of perhaps 3 Gyr for the ''old EROs", corresponding to formation 
at $z \approx 2.5$ (Cimatti et al., 2002a). 
Near $z=1$, the old E's are much more strongly clustered 
(Daddi et al., 2002). From their statistics for 480 K-band selected objects, 
with redshifts up to 1.5 or more, Cimatti et al. (2002b) attempt to test current 
galaxy formation models, and favor the pure luminosity evolution of old galaxies 
against hierarchical merging evolution. 

The Fundamental Plane is still ''fundamental'' at large $z$.  The small scatter 
observed in the M/L ratio derived from the FP imposes
tight constraints on the dynamics,  IMF and ages of the
stellar populations (Renzini \& Ciotti 1993). 
J$\phi$rgensen et al. (1999) have obtained photometric and kinematic data and 
derived the FP for two clusters near $z=0.18$. 
They compare their results with those of van Dokkum \& Franx (1996), and 
Kelson et al. (1997) for still more distant clusters, and find a 
smooth decrease of M/L with $z$. The results are consistent with passive 
evolution 
of a stellar population formed at $z \approx 5$ (for $q_0=0.5$). 
This agrees with the conclusions reached by using other clues of evolution, i.e. 
the Kormendy relation for Ziegler et al. (1999), and the colour or line-indices
relations.  The work of Pahre et al. (2001) again demonstrates 
that the SB intercept of the FP, the rest frame U-V colour and the absorption 
line strengths, all evolve passively  in the range $0 < z < 0.6$ covered 
by the data. 

The small scatter 
observed in the Mg$_2$ - $\sigma_d$ (or colour - $\sigma_d$) relation suggests 
that the bulk of  stars of luminous ellipticals must be rather old, and formed 
in a short time interval of the order of $2-3$ Gyr. 
In the evolutionary models by Idiart et al. (2002b), the more massive galaxies
have a higher star formation efficiency and therefore the slope of the
relation Mg$_2$ - $\sigma_d$ increases slightly with the redshift. In Fig. 10
the plot of this relation for different redshifts clearly demonstrates this
effect.


Mg$_b$ indices of E's in three clusters at intermediate 
redshifts (z $\sim$ 0.37) were compared to those in nearby (z $\sim$ 0) clusters 
by Ziegler \& Bender (1997). These authors found a mean difference 
$\Delta Mg_b = <(Mg_b)_{z=0}-(Mg_b)_{z=0.37}> \approx 0.4$ \AA \,\  between both 
samples, compatible with a "passive" evolution. In fact, using their data on
the velocity dispersion and theoretical  
Mg$_2$ - $\sigma_d$ relations, 
the predicted mean difference $\Delta Mg_b$ is 0.34 \AA (Mg$_b \approx 15Mg_2$), 
consistent with observations.
Moreover a brightening of about -0.5 mag at fixed $\sigma_d$ is observed in 
the B-magnitude, also consistent with the theoretically expected increase of 
about -0.44 mag in the V filter. 

The $z$ variation of the CM diagram in clusters has been studied up to $z 
\approx 1$ by 
Stanford et al. (1995, 1998). The results are again explained by the passive 
evolution of a population formed  at $z > 2.5$. van Dokkum et al. (2000) have 
obtained extensive color data for E's in a  cluster at $z=0.83$, concluding by a 
passive evolution of a population mix older 
than $z \approx 2.5$. 

The r\^ole played by the $\alpha$-elements, represented mainly by 
magnesium, in the evolution of early-type galaxies was recently reviewed by 
Thomas et al. (2002). They use SSP models with variable [$\alpha$/Fe] 
ratios to analyze line-indices in a sample of 126 field and cluster local 
E-galaxies, and find a correlation  
between the $[\alpha/Fe]$ ratio and the velocity dispersion (see also 
de Freitas Pacheco 1996), and also between this ratio and the estimated mean 
population age.  They conclude that  the more massive the galaxy, the shorter 
is its star formation scale, and the higher is the redshift of the bulk of the 
star 
formation. These findings are not compatible with the predictions of 
hierarchical 
models of galaxy formation. These conclusions are in line with 
evolutionary models as presented by Idiart et al (2000b).   

The morphological classification of galaxies, extended to large $z$ from HST 
images,  
leads to rather ambiguous results. This is in part due to the intrinsic 
difficulty  
of this exercise, with different observers obtaining divergent results from the 
same 
data (Fabricant  et al. 2000). Dressler et al. (1997) found a relative high 
fraction of spirals accompanied by a low fraction of S0's in  
clusters within the redshift range $0.3 < z < 0.5$, concluding that the 
former classes transform into the later morphological types. Other authors 
(see, for instance, Andreon et al. 1997) find a much "milder" evolution of the 
S/S0 ratio. van Dokkum et al. (2000) studied the rich cluster 
MS1054-03 at $z = 0.83$ and found a deficit in the population of early-type 
galaxies, reaching a factor of 2 from $z = 0$ to $z  \sim 1.0$. Moreover, they  
have estimated that the fraction of very close
binary (or interacting) galaxies, probably  future mergers, is 17\%, 
which could support the idea that ellipticals are ''merged" spirals.

In conclusion, the stars in ellipticals (or most of them by mass and light) 
are no doubt old from their spectral properties and the corresponding $z$ 
variations. 
But they may have been reassembled much later into their present nice
spheroidal systems: Franx \& van Dokkum (2001) have proposed a mechanism to 
reconcile
these conflicting indications, at least as far as cluster galaxies are 
concerned.

\section{Concluding remarks}
\vspace{-0.25 cm}
The recent high quality imaging and kinematical data on ellipticals have
shown that besides tri-axiallity, these galaxies have a complex structure
including the presence of inner disks, cores and cuspy light profiles.
The tri-axiallity inferred from radial variations of the axial ratios
was further supported by new data, suggesting substantial rotation
along the minor axis. Moreover, such high quality data demonstrated
the existence of small central components not sharing the rotation
of the main body, which are interpreted as the result of capture of
a small nearby galaxy.

There is presently  strong evidences that most (if not all) ellipticals 
host in their centers a massive black hole able to modify the profile
of the very central matter distribution (baryonic and non-baryonic). These
SMBH form perhaps contemporaneously with all stellar spheroids, a picture
supported by the observed correlation between the SMBH mass and the
central velocity dispersion of the galaxy.

The dynamical evidence for the presence of dark matter in the central
regions either of spirals or ellipticals is not conclusive, since
the predicted "cuspy" profiles from numerical simulations are not
seen in rotation curves, and stars can provide the required gravitational
field inside the optical radius to explain kinematical data. However,
at larger distances, the confinement of the hot gas seen in X-rays implies
the existence of a substantial amount of dark matter in halos, also
supported by recent kinematic data on planetary nebulae 
and globular clusters, in the outskirts of giant ellipticals. 
The extensive numerical simulations performed by several authors indicate
that the density distribution inside dark matter halos, formed through
successive merging episodes, can be represented by an "universal"
profile, independent of the parameters defining the cosmological model,
the mass of the halo or the redshift.

The understanding of the heating and cooling of the baryon component is
an essential key in the process of galaxy formation, since it permits
the definition of a critical mass (for a given dark halo mass) above which
star formation is triggered. However many problems still remain unsolved, 
as the origin of the angular momentum and its relation with the
galactic morphology. In the hierarchical scenario, the morphology is
the consequence of a particular merger and accretion history that
galaxies experience during their assembly. Disks are the result of
gas accreted smoothly from the intergalactic medium, while spheroids
would be formed through major merger events where disks are mixed
and relaxed violently. 

The bulk of the stellar populations of E's is quite old (12-13 Gyr),
according to the interpretation of the integrated colors and
the color-magnitude diagram. However, based mainly on the H$\beta$ and  $<Fe>$ 
indices, 
some authors found galaxies with stellar populations
substantially younger, sometimes in contradiction with the observed
colors. It should be emphasize that other effects than age or emission may
affect the $H\beta$ index as, for instance, the horizontal branch
morphology (de Freitas Pacheco \& Barbuy 1995; de Freitas Pacheco 1998).
Recent star formation episodes triggered by accretion of
intergalactic material or small nearby galaxies cannot be excluded, and appears 
prominent in several local objects. A further indication of an initial and 
intense 
star formation
period is provided by the relative abundance of the $\alpha$-elements, which
is a "clock" suggesting that the majority of the stars were formed in
a timescale of 2-3 Gyr. Most of the E's have mean metallicities near solar, 
which
is an indication of a population mix, constituted of different generations
required to the build up of the chemical elements.

The colors, the color-magnitude diagram (most probably a metallicity-mass 
diagram)
and the relative abundance of the $\alpha$-elements favors an evolutive
scenario where the star formation activity begins after the assembly of a 
critical
mass of gas, according to the description given previously. After the 
triggering,
most of the stars are formed in a relatively short timescale (2-3 Gyr) and then
evolve "passively". These conclusions, as seen in Section 5.2, are reinforced
by the small scatter in the M/L ratio derived from the Fundamental Plane and its
decrease with redshift, equivalent to a brightening of the
galaxies with $z$. The evolution of the luminosity function of "red" galaxies
also points in such a direction, as well as the evolution of the Mg$_b$ - 
$\sigma_d$
relation, although the available data on the latter cannot be considered as
conclusive. On the other hand, the main argument in favor of an assembly of
spirals to form ellipticals is the possible decrease of the fraction of the
latter at higher redshifts, an evidence criticized by some authors on the
basis of the difficulties to distinguish morphological types at high $z$. 

The two scenarii, ''monolithic" versus ''hierarchical", sketched in our 
introductory section to describe the origin and evolution of ellipticals, 
will probably be no more tenable in a near future, at least in their simplistic 
forms. Recently observers and theoreticians have made great strides towards a 
synthesis. 
The epochs of major stellar formation have been identified from very deep 
multi-wave-length surveys (Steidel, 1999) and from the analysis of 
the Cosmic IR Background (Gispert et al. 2000). 
The stellar formation rate strongly declined after $z=1$ and was possibly 
maximal 
near $z=2-3$. The brightest galaxies detected at such epochs 
(Lyman Break Galaxies) seem to be analogous 
to the local luminous IR galaxies with intense star formation behind an heavy 
dust 
''curtain". Guiderdoni (2002) summarizes the work undertaken to model, {\em 
within a supercluster-size volume}, the 
formation and hierarchical evolution of these objects, which could be 
the progenitors of massive present day galaxies, both spirals and ellipticals. 

We thank U. Frisch and S. Matarrese for helpful comments. R.M.
is supported by European TMR network (contract HPRN-CT-2000-00162).     

{}
\newpage

\begin{figure}[hbt]
\centerline{
        \epsfysize=\royalength
        \epsfbox{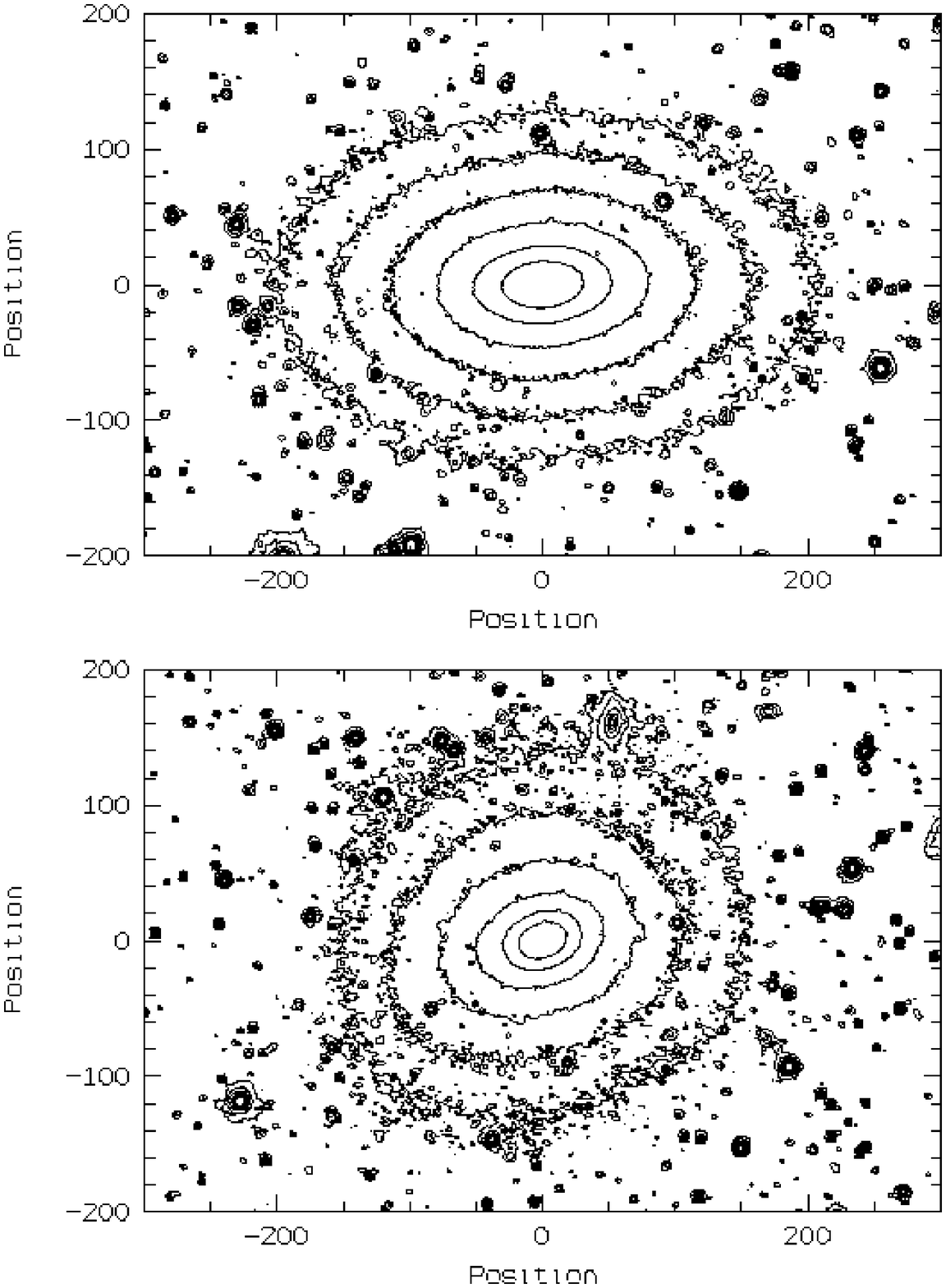}
           }
\vspace{1.2cm}
\caption
{
Isophotal maps of symmetric and asymmetric E-galaxies: isophotes are
shown for V magnitudes per arcsec square, from 20.5 up to 25.5 (steps of one
magnitude) and positions are given with respect to the galaxy center.. Frames 
were 
obtained at Haute Provence Observatory (Idiart et al. 2002a). Upper panel: NGC 
4473,
a diE in Virgo and lower panel, NGC 5982, a boE in a small group. The latter 
shows
important asymmetries as  large ellipticity variations, isophotes twist and 
drift
of the center.
}
\label{1}
\end{figure}
\begin{figure}[hbt]
\centerline{
        \vspace{-1.0cm}
        \epsfysize=3.4 true in
        \epsfbox{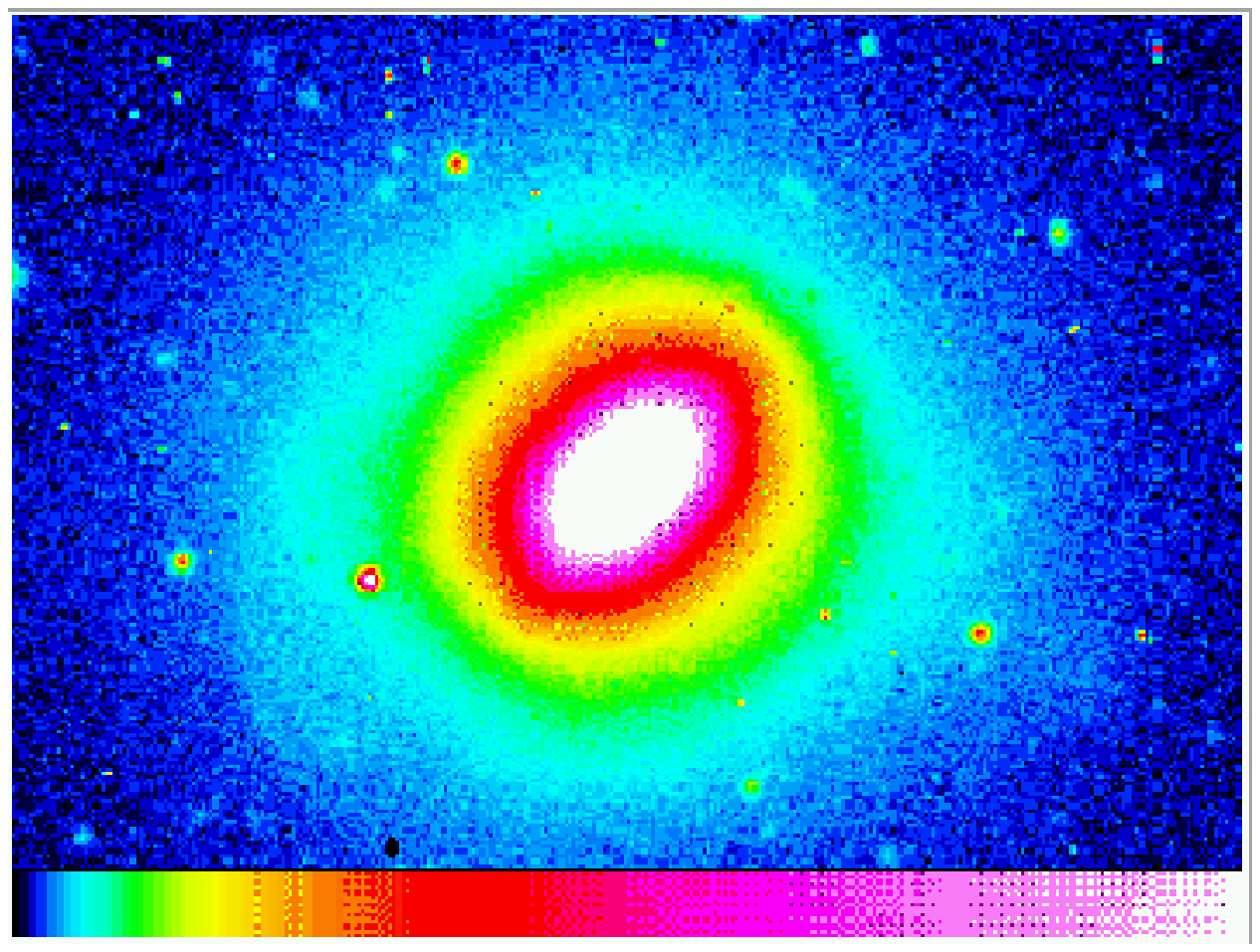}
           }
\label{2a}
\end{figure}
\begin{figure}[hbt]
\centerline{
        \epsfysize=3.4 true in
        \epsfbox{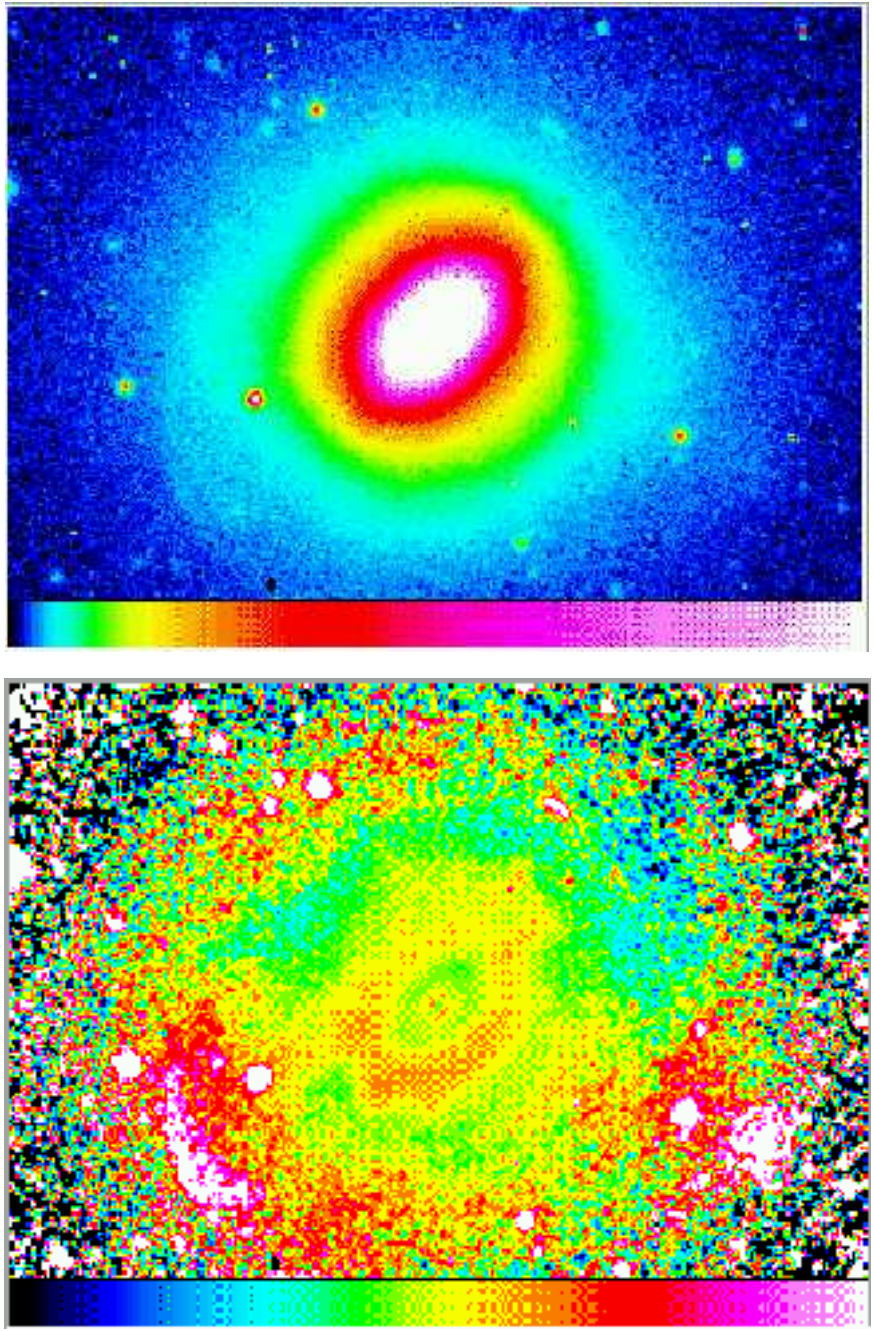}
           }
\caption
{
Peculiarities in the NGC 3610; upper panel shows a V band image and 
the
lower panel shows a map of brightness fluctuations where radial variations were
cancelled by an ad hoc mask. Notice a SE sharp bright feature ("shell" or 
"ripple" 
in the usual terminology); another bright feature appears in the SW.
}
\label{2b}
\end{figure}
\begin{figure}[hbt]
\centerline{
        \vspace{-0.8cm}
        \epsfysize=\royalength
        \epsfbox{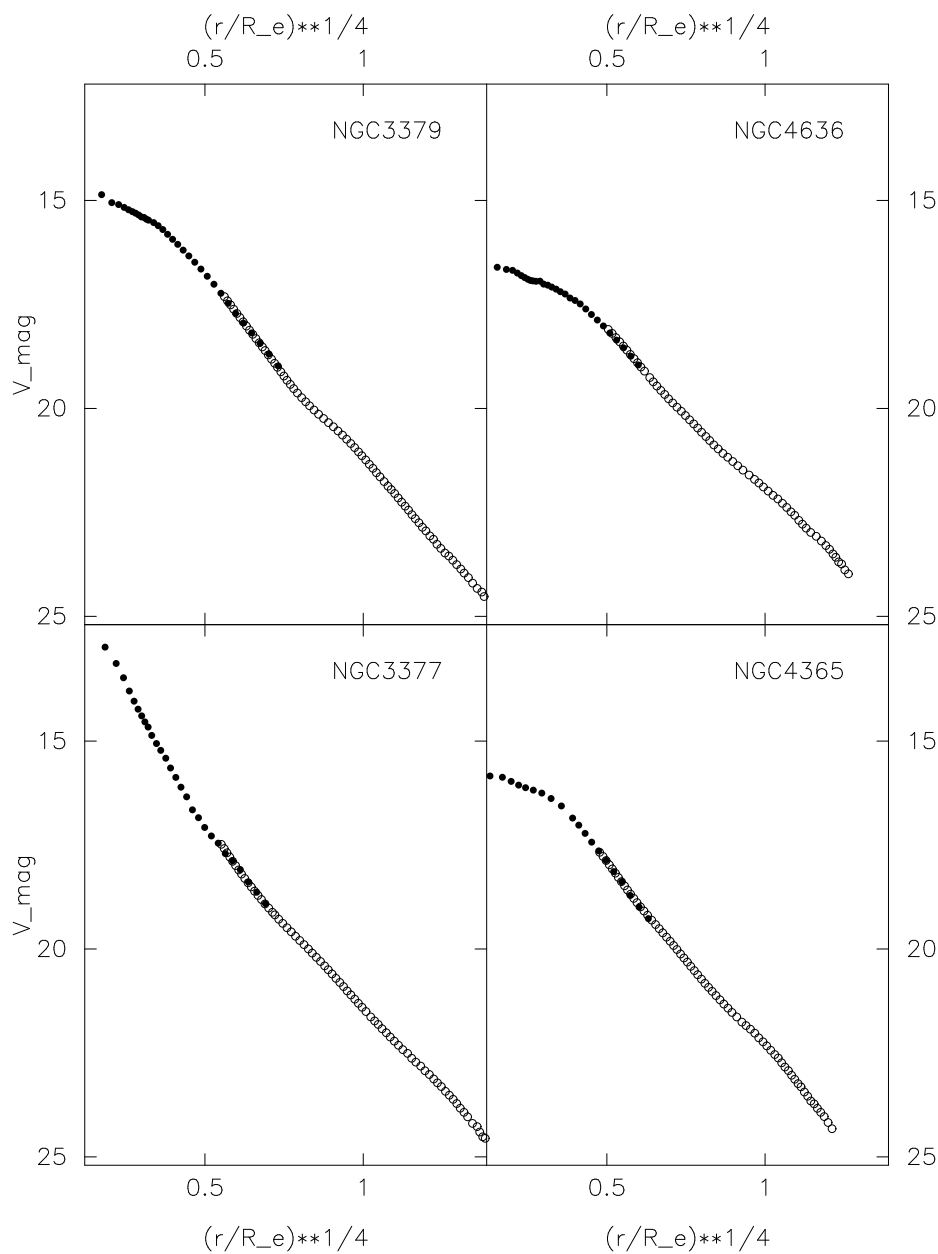}
           }
\vspace{1.0cm}
\caption
{
Brightness profiles in V light for different E-type galaxies. Filled
circles are HST data by Lauer et al. (1995); Carollo et al (1997a); Gebhardt et
al. (2000b), whereas open circles are wide field data by Idiart et al. (2000a).
Abscissae are normalized with respect to the effective radius, so homologous
profiles appear as parallel lines. Notice that NGC 3377 has a "cuspy" profile
running well above the "r$^{1/4}$-law" whereas "core-like" profiles can be seen
in the two boE giants NGC 4636 and NGC 4365.
}
\label{3}
\end{figure}
\begin{figure}[hbt]
\centerline{
        \vspace{-0.8cm}
        \epsfysize=\royalength
        \epsfbox{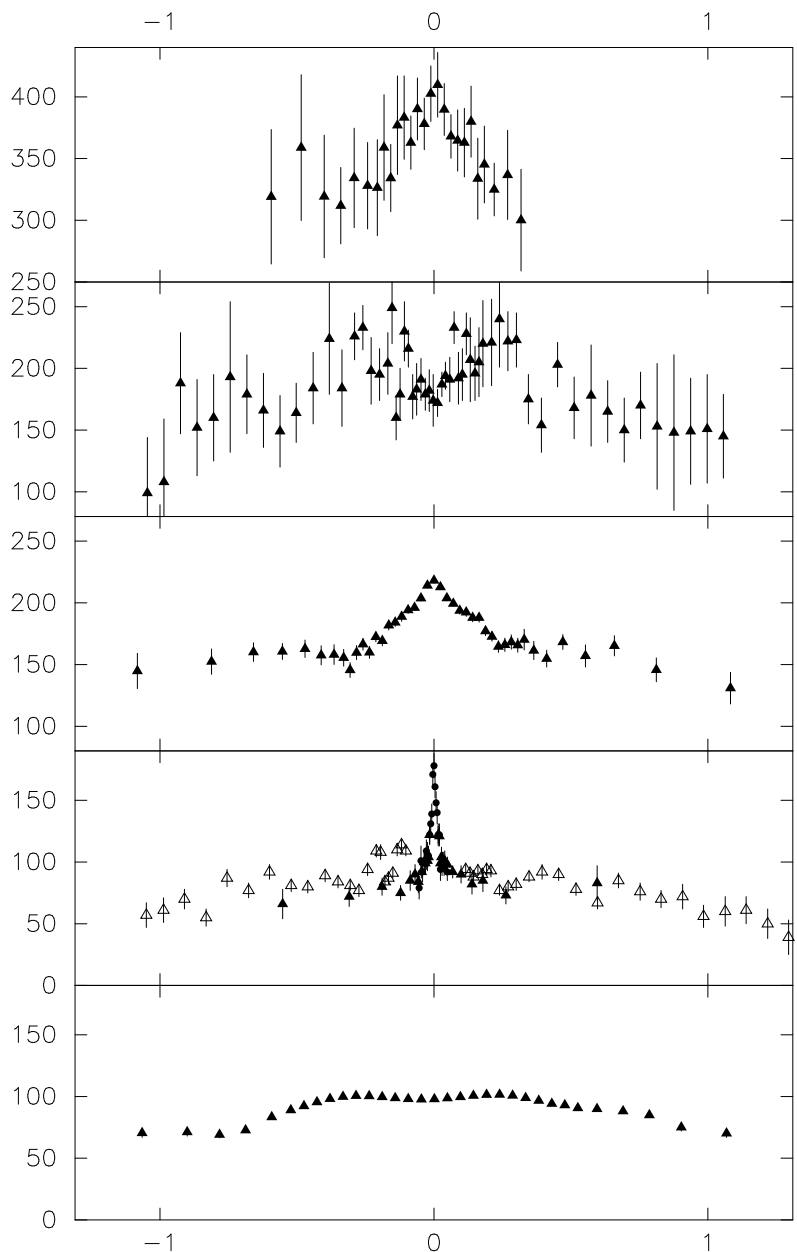}
           }
\vspace{1.cm}
\caption
{
Examples of velocity dispersion profiles in ellipticals of different
luminosities. From top to bottom: NGC 4889 a cD galaxy in Coma, having one of 
the
largest $\sigma_d$ ever recorded (Mehlert et al. 2000); NGC 2768 with an 
exceptional 
central minimum (Simien \& Prugniel 1997); NGC 3379 a prototype profile (Statler
\& Smecker-Hane 1999); NGC 3377, a faint object (diE) with a sharp central peak, 
interpreted
as an evidence for the presence of a central SMBH (Kormendy et al. 1998; Simien
\& Prugniel 2002); NGC 4387, a faint boE satellite of a pair of giants.
}
\label{4}
\end{figure}

\begin{figure}[hbt]
\centerline{
        \vspace{-0.8cm}
        \epsfysize=\royalength
        \epsfbox{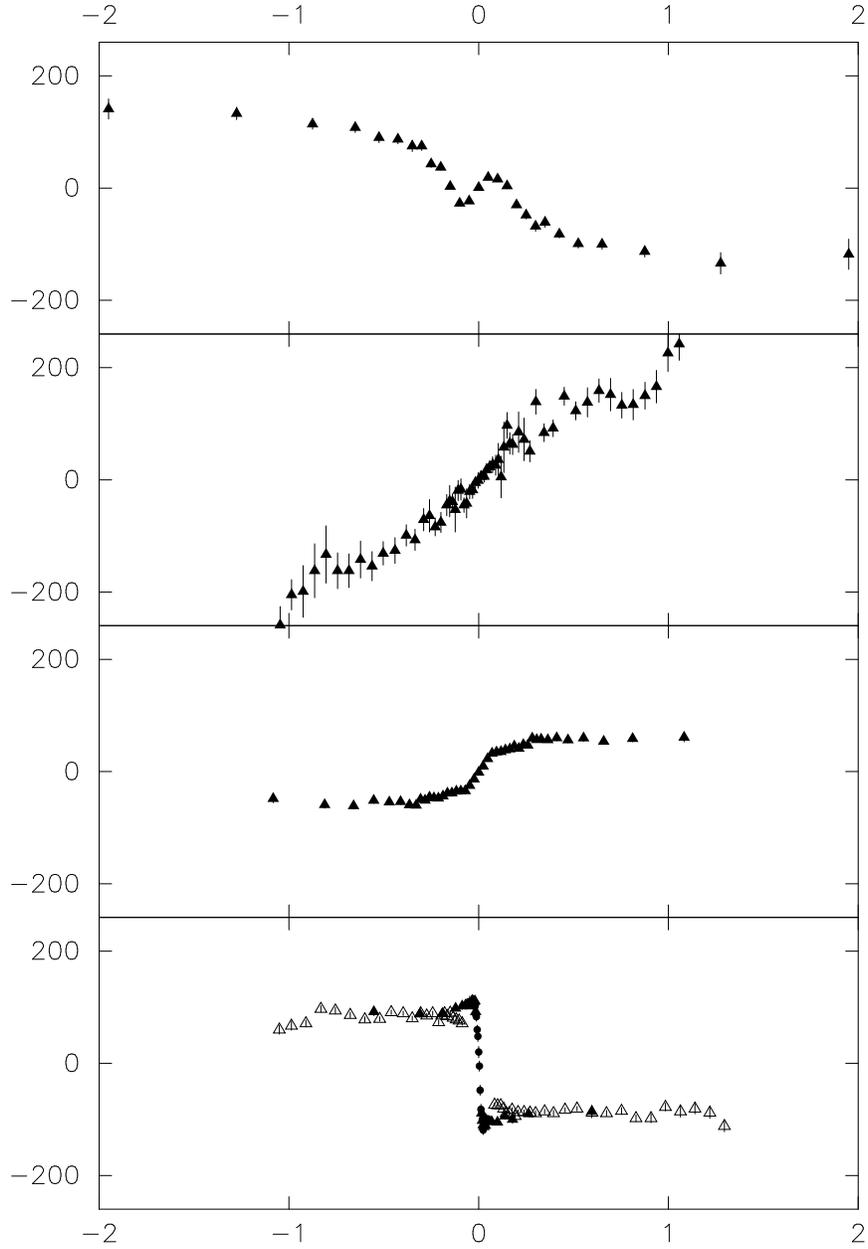}
           }
\vspace{1.0cm}
\caption
{
Examples of rotation profiles in ellipticals. From top to bottom:
NGC 1700, displaying an outer "twist" of the rotation axis (Statler et al. 
1996);
NGC 2768, shows an increasing velocity up to the last data points; NGC 3379,
a galaxy with a modest apparent rotation due to projection effects; NGC 3377, a 
diE
with fast rotation in the central region.
}
\label{5}
\end{figure}
\begin{figure}[hbt]
\centerline{
        \vspace{-0.6cm}
        \epsfysize=4 true in
        \epsfbox{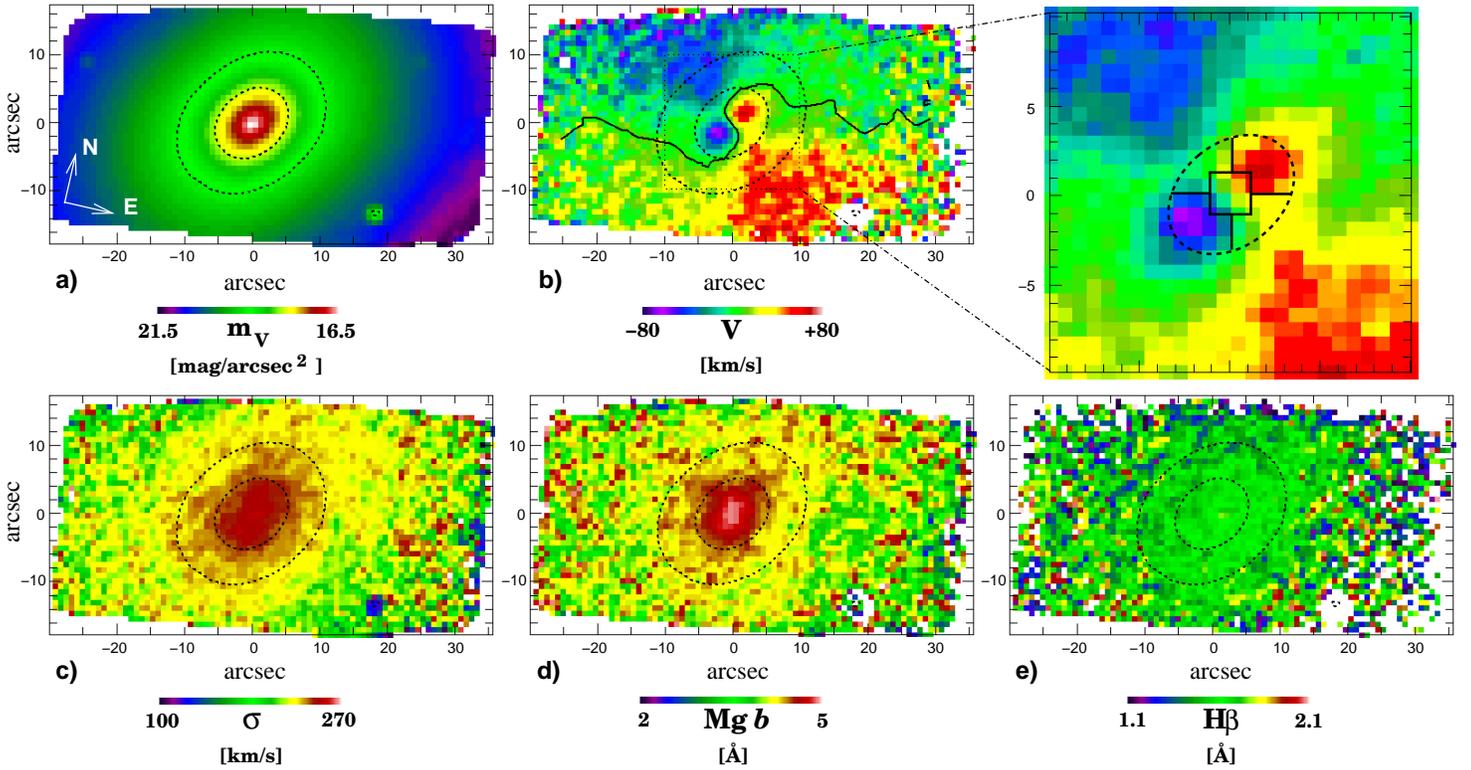}
           }
\vspace{1.0cm}
\caption
{
3D maps of kinematical parameters of NGC 4365 from SAURON. Upper 
panels: left,
V magnitudes; center, line of sight velocities with details at the right panel.
Lower panels: left, velocity dispersion; center, Mg$_b$ index; right, H$\beta$ 
index.
Reproduced from Davies et al. (2001) by courtesy of the authors and of
the American Astronomical Society.
}
\label{6}
\end{figure}
\begin{figure}[hbt]
\centerline{
        \vspace{-0.8cm}
        \epsfysize=8 true in
        \epsfbox{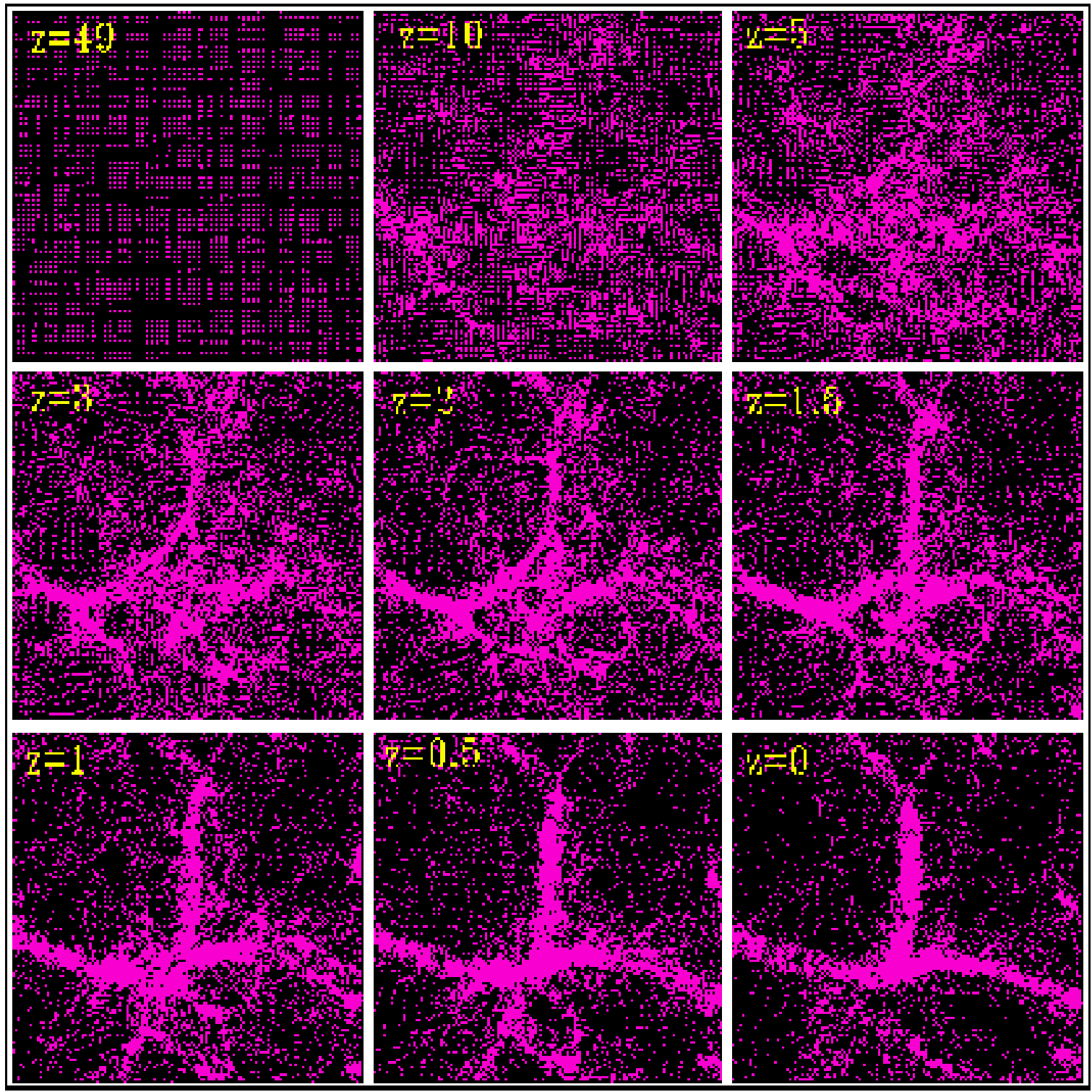}
           }
\vspace{1.0cm}
\caption
{
The distribution of DM particles are shown at various redshifts. The 
plots are thin
slices of sides 32x32x20 Mpc, taken from runs of a $\Lambda$CDM model and using
a P$^3M$ code, including 128$^3$ particles in a box size of 64h$^{-1}$ Mpc. The 
evolution demonstrates the formation, under gravitational instability, of 
pancakes,
filaments, nodes and voids, from a quasi-homogeneous initial condition.  
}
\label{8}
\end{figure}

\begin{figure}[hbt]
\centerline{
        \vspace{-0.8cm}
        \epsfysize=7.5 true in
        \epsfbox{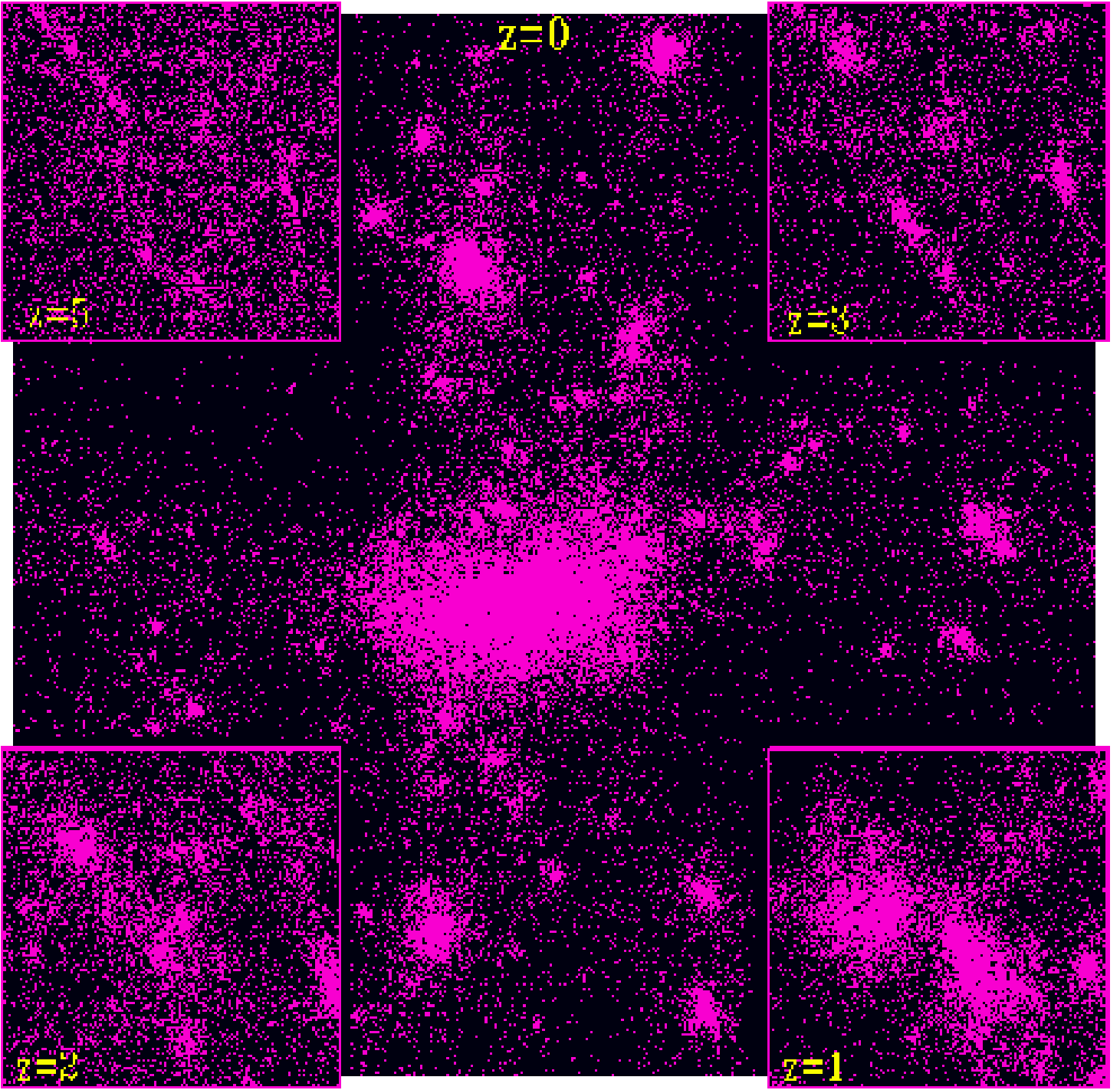}
           }
\vspace{1.0cm}
\caption
{
The results of the same simulation are shown at much smaller scales 
(slices
of dimensions 5x4x20 Mpc). At such scales DM halos form and grow by merger and 
accretion.
The plots show a growth of two halos and their eventual merger leading to a 
single
halo at the present time.
}
\label{9}
\end{figure}
\begin{figure}[hbt]
\centerline{
        \vspace{-0.8cm}
        \epsfysize=\royalength
        \epsfbox{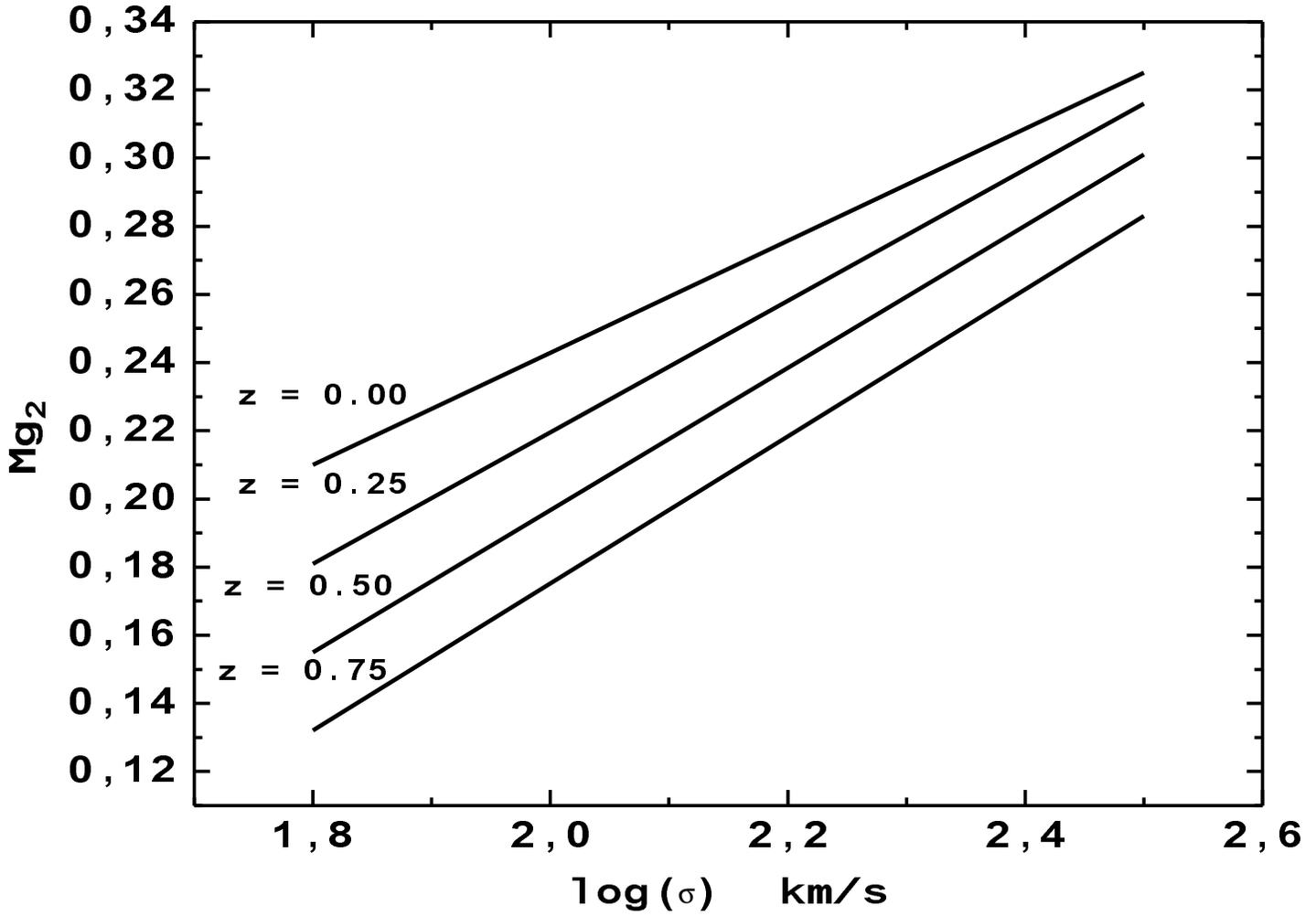}
           }
\vspace{1.0cm}
\caption
{
Plot of theoretical Mg$_2$ indices versus the central velocity 
dispersion
for different redshifts (z = 0.0, 0.25, 0.50 and 0.75). Notice the increasing 
slope at 
higher redshifts (respectively equal to 0.164, 0.192, 0.208 and 0.216), 
resulting
mainly from the higher star formation efficiency in more massive galaxies.
}
\label{10}
\end{figure}

\end{document}